# Particle-like topologies in light

Danica Sugic[1,2,3*], Ramon Droop[4*], Eileen Otte[4], Daniel Ehrmanntraut[4],
Franco Nori[3,5], Janne Ruostekoski[6], Cornelia Denz[4], Mark R. Dennis[1,2,7*]

[1] School of Physics and Astronomy, University of Birmingham, Birmingham B15 2TT, UK

[2] H H Wills Physics Laboratory, University of Bristol, Bristol BS8 1TL, UK

[3] Theoretical Quantum Physics Laboratory, RIKEN Cluster for Pioneering Research, Wako-shi, Saitama 351-0198, Japan

[4] Institute of Applied Physics and Center for Nonlinear Science (CeNoS), University of Muenster, 48149 Muenster, Germany

[5] Physics Department, University of Michigan, Ann Arbor, Michigan 48109-1040, USA

[6] Physics Department, Lancaster University, Lancaster LA1 4YB, UK

[7] EPSRC Centre for Doctoral Training in Topological Design, University of Birmingham, Birmingham B15 2TT, UK

**Three-dimensional (3D) topological states resemble truly localised, particle-like objects in physical space. Among the richest such structures are 3D skyrmions and hopfions[1,2,3], that realise integer topological numbers in their configuration via homotopic mappings from real space to the hypersphere[4] (sphere in 4D space) or the 2D sphere. They have received tremendous attention as exotic textures in particle physics[1,2,5], cosmology[6,7], superfluids[3,8], and many other systems[2,9,10]. Although a small number of these 3D states have been realised in different laboratory experiments[10,11,12], a full characterisation of topological properties presents extreme challenges, and attention has focused on much simpler, planar analogues[13,14,15,16,17,18]. Here we experimentally create and measure a topological 3D skyrmionic hopfion in fully structured light. By simultaneously tailoring the polarization and phase profile, our beam establishes the skyrmionic mapping by realising every possible optical state in the propagation volume. The resulting light field's Stokes parameters and phase are synthesised into a Hopf fibration[19] texture. We perform volumetric full-field reconstruction of the $\Pi_3$ mapping, measuring a quantised topological charge, or Skyrme number, of 0.945. Such topological state control opens new avenues for 3D optical data encoding and metrology. The Hopf characterisation of the optical hypersphere endows a new perspective to topological optics, offering experimentally-accessible photonic analogues to the gamut of particle-like 3D topological textures, from condensed matter to high-energy physics.**

Nontrivial 3D topology has inspired many descriptions of fundamental particles. Motivated by Lord Kelvin's knotted vortex atom hypothesis[20], Tony Skyrme[1] in 1961 proposed a topological model for nuclei: particle-like continuous fields in 3D space now called skyrmions. These map 3D real space to the hypersphere (i.e. the unit sphere in four dimensions, also known as the 3-sphere[4]), parametrising the field. The skyrmion configuration wraps around the hypersphere an integer number of times called the Skyrme number. Skyrmions are now seen as a particular example of more general 3D topological solitons[2,3,7], related to other topological textures such as monopoles and hopfions – the latter being fields with a 2-sphere parameter space (i.e. unit sphere in three dimensions). 3D topological textures have been studied theoretically as hypothetical objects in various systems, including high-energy physics[2,5], condensed matter[3,7,8,9], and early-universe cosmology[6]. In recent years, 3D skyrmions and hopfions have been experimentally realised in cold quantum matter[11,12] and liquid crystals[10].

So-called "baby skyrmions" are the two-dimensional (2D) counterpart of 3D skyrmions: fields in 2D physical space which map to, and wrap around, a 2-sphere parameter space. Their study is much more developed in theory and experiments, notably in non-singular superfluid vortices[13] including those imprinted by structured light[14], and especially magnetic systems[15]. Here the direction of spin at each point provides the 2-sphere parameter space, and magnetic skyrmion excitations have the potential to represent topological bits for low-power computer memory and processing[15]. Recently, 2D baby skyrmion configurations were created in optical systems, as the direction of electric field vectors, or photon spin, near a material interface[16,17], displaying dynamics similar to magnetic skyrmions[18]. In propagating laser light, optical polarization can be structured into full Poincaré beams[21], which realise every state of elliptic polarization in the transverse plane. These beams can also be interpreted as 2D baby skyrmions[22], since the Poincaré sphere, as the 2-sphere parameter space, parametrises transverse, elliptic polarization states. However, 3D particle-like topological objects have not been considered either theoretically or experimentally in optical fields.

Optical realisations of 3D topological states can take various forms. Much interest has focused on singularity lines, such as optical vortices or polarization singularities (e.g. C lines)[22]. In structured light, with amplitude, phase, and polarization spatially varying, these can be woven into loops, links and knots[23,24] and organise Möbius strips[25]. The state of elliptic polarization is right- or left-handed circular (RH, LH) on C lines, often described as a skeleton of the complex optical polarization field[26]. Topologically structured light has a wide range of applications including enhanced free-space optical communications[27] and advanced trapping[28], and is related to optical currents[29] and orbital angular momentum[30].

Singular lines are topologically characterised by the fundamental homotopy group $\Pi_1$. The homotopy group $\Pi_3$, on the other hand, defines topological particles such as 3D hopfions and skyrmions[2]. It is natural to ask whether these 3D excitations can be created in structured light. We describe here the design, generation and measurement of a structured, propagating beam of laser light realising such a mapping, unifying particle-like 3D topologies in free space optics with those studied in high-energy physics, cosmology and various kinds of condensed matter.

## The optical hypersphere of polarization and phase

Spatially extended polarized light is represented by a complex transverse electric field vector at each point $\boldsymbol{r}$ in the propagating beam. Its RH and LH components are represented by the complex-valued scalar functions $E_R(\boldsymbol{r})$ and $E_L(\boldsymbol{r})$, and the pair $(E_R, E_L)$ which characterises the optical state at each point is assumed normalized, i.e.

$$(\mathrm{Re}E_R)^2 + (\mathrm{Im}E_R)^2 + (\mathrm{Re}E_L)^2 + (\mathrm{Im}E_L)^2 = 1.$$

Therefore, this normalized optical field defines a mapping from each point in 3D real space to a point on the 3-sphere, which we call the *optical hypersphere*. The optical hypersphere is conveniently parametrised using spinorial angles $\alpha, \beta, \gamma$:

$$E_R = \cos\frac{\beta}{2}\mathrm{e}^{\mathrm{i}(\gamma-\alpha)/2} \quad \text{and} \quad E_L = \sin\frac{\beta}{2}\mathrm{e}^{\mathrm{i}(\gamma+\alpha)/2}, \tag{1}$$

for $0 \leq \beta \leq \pi$, $-\pi < \alpha \leq \pi$ and $-2\pi < \gamma \leq 2\pi$. The angles $\alpha, \beta, \gamma$ have a direct interpretation in terms of the polarization and phase of the electric field state: with $S_1, S_2, S_3$ the normalized Stokes parameters, $\alpha = \arctan(S_1, S_2)$ is the polarization azimuth, and $\cos\beta = S_3$ is the polarization ellipticity; $\gamma = \arg E_R + \arg E_L$ is the sum of the two electric field components' phases[26,31]. Further details of these parameters and their relationship with the hypersphere and the Poincaré sphere (2-sphere) parametrising polarization may be found in the Supplementary Information.

The full Poincaré sphere of polarization states can be realised in a transverse plane of a structured light field, created from the superposition of two, differently structured, LH and RH beam components, similar to a full Poincaré beam[21]. At each spatial point, the optical field has some elliptical polarization state characterised by $\alpha, \beta$. In 3D, points of constant elliptical polarization lie on filaments, generalizing RH and LH circular polarized C lines. 3D real space is filled by the set of polarization filaments, constituting a polarization texture (Fig. 1a). Each filament corresponds to a point on the Poincaré sphere (Fig. 1b), and many filaments cross each plane (Fig. 1c). Although the polarization is fixed on the filaments, the optical phase smoothly varies along them (Fig. 1c insets). Any 3D structured light field with varying transverse polarization can be represented by such a texture.

The 3-sphere supports the Hopf fibration[19], a fibre bundle which divides it into linked circles. In the optical hypersphere, each fixed polarization state (with $\alpha, \beta$ constant) traces out a circle as the phase $\gamma$ goes through a $4\pi$ cycle. The phase and polarization parameters therefore realise the Hopf fibration in the optical hypersphere (this is explained in detail in the Supplementary Information). The Poincaré sphere is interpreted here as the base space of the fibration[31]. We design a 3D structured beam which realises all the transverse states of light, including polarization and phase, in its focal volume (real space). It displays the 3D Hopf fibration topology in a configuration we call a skyrmionic hopfion. The skyrmionic hopfion realises, in real space, an image of the Hopf fibration in the optical hypersphere. The fixed polarization filaments can be represented as a 3D topological texture of entwined curves, in which each pair of loops is linked.

**Experimentally realising the skyrmionic hopfion**

We design the skyrmionic hopfion structure in light by superimposing carefully chosen combinations of vectorial Laguerre-Gauss beams $\mathrm{LG}_{\ell,p}$[23,30]. The LH component, $E_L$, is chosen to be the Laguerre-Gauss beam $\mathrm{LG}_{-1,0}$, with a negative-signed optical vortex along the beam axis[23]. The RH component, $E_R$, is chosen as a superposition of the Laguerre-Gauss beams $\mathrm{LG}_{0,0}$ and $\mathrm{LG}_{0,1}$, with a circular vortex loop in the focal plane centred on the axis[23]. Therefore, the net polarization field has a RH C line along the axis, threading a LH C line loop in the focal plane. The C lines, at which $\beta = 0, \pi$, organise the rest of the texture: between them are nested tori with $\beta =$ constant, including the particular L surface of linear polarization at $\beta = \pi/2$, analogous to vortices in other skyrmionic textures[8]. Details of the superposition optimisation are given in the Methods and Supplementary Information.

Experimentally, the RH and LH beam components are separately shaped by a spatial light modulator (SLM) (Fig. 2), before being combined in a joint beam path to shape the skyrmionic hopfion (Supplementary Fig. 4). The total polarization state and phase of the resulting beam are measured at each point in the propagating volume via vectorial full-field reconstruction (VFFR, see Methods). For the VFFR, we combine established metrological techniques; namely, Stokes polarimetry, interferometry, and digital propagation[32]. Our approach explicitly relates measurements in different 2D planes, reconstructing the full 3D field volume. Further details of the experiment are given in the Methods and Supplementary Information.

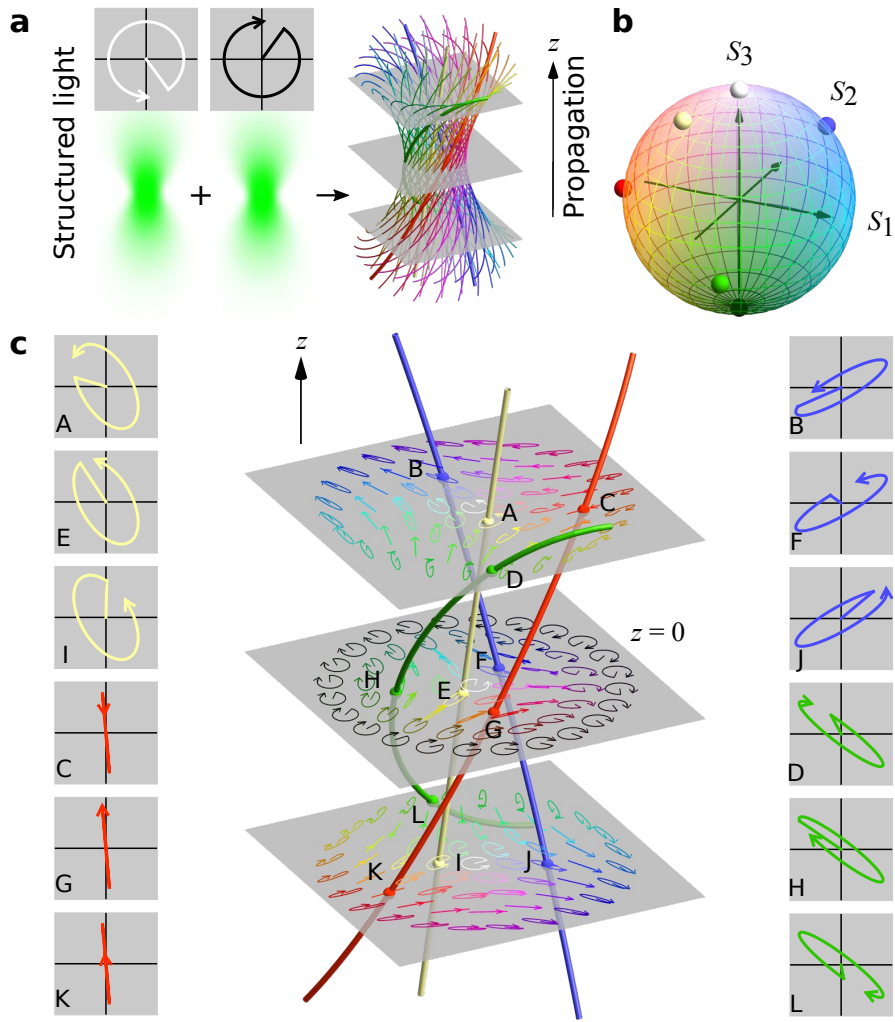

**Figure 1: 3D Optical Polarization Texture.** A light field with position-dependent transverse polarization and phase is created in a volume from the superposition of RH and LH circularly polarized beams, whose amplitude and phases are carefully structured. Spatial points characterised by the same state of elliptic polarization lie on filaments (**a**). The 3D polarization texture can be visualised by colouring the filaments according to the position of its polarization ellipse on the Poincaré sphere (**b**). The azimuthal angle $\alpha$, representing the ellipse orientation, is coloured with the hues and the polar angle $\beta$, representing the ellipticity, is associated with the saturation levels. The sphere's poles, representing the circular polarized states, are black (LH) and white (RH). Each optical state also has a phase, represented by the position of the arrow along the polarization ellipse. In the transverse plane in (**c**), states of light are fully described by colours and arrowed ellipses. Along filaments of constant polarization, the phase on the ellipses varies smoothly, as shown in the insets for three representative planes.

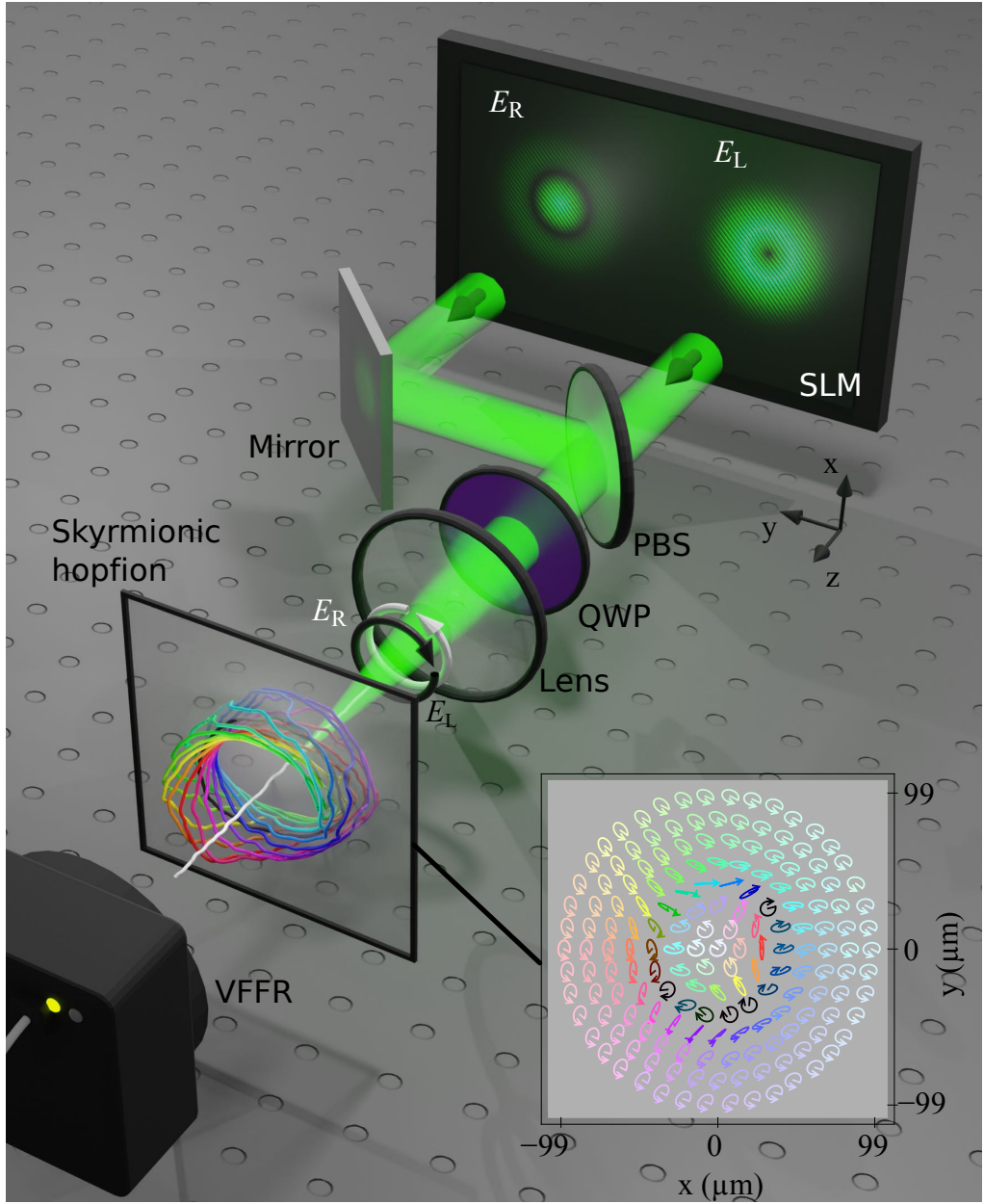

**Figure 2. Sketch of experimental setup.** Beams $E_R$ and $E_L$ are generated on a spatial light modulator (SLM) and superimposed on axis by a polarizing beam splitter (PBS). A quarter wave plate (QWP) transforms $E_L$ into left circular (black) and $E_R$ into right circular (white) polarization. Around the focal spot of the lens, the skyrmionic hopfion appears in a cuboid of size $99.4\,\mu m \times 99.4\,\mu m \times 26.6\,mm$. The inset shows the polarization texture in the focal plane (colour coded as the Poincaré sphere in Fig. 1b), consistent with theory in Fig. 1c. Measurements of amplitude, phase and polarization are enabled by volumetric full field reconstruction (VFFR, see Methods).

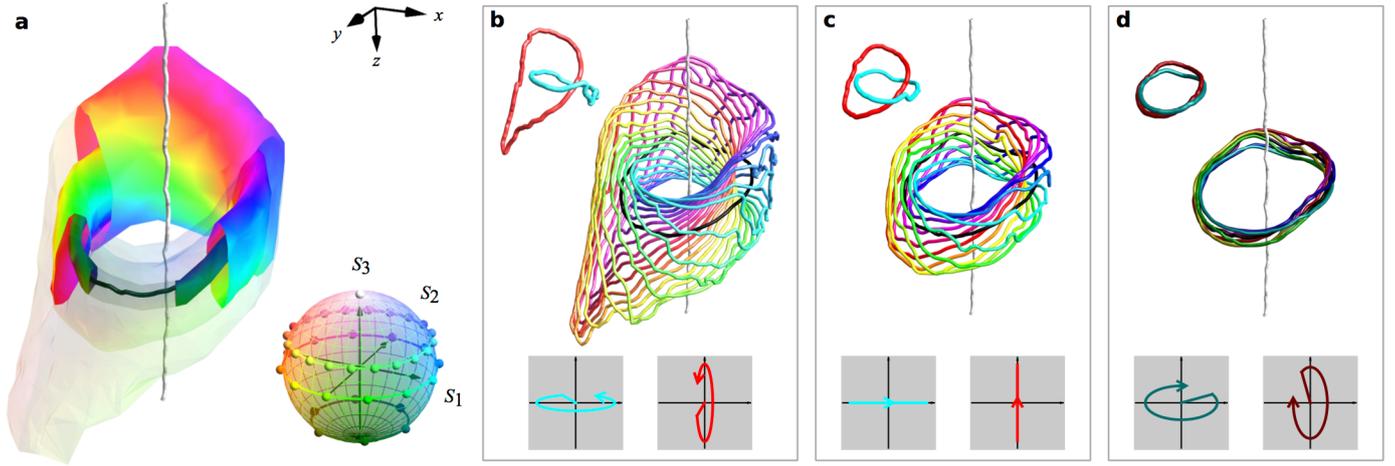

**Figure 3. Visualising the topology of the focal volume.** The optical texture is reconstructed from the polarization and phase measurements via the VFFR of the optical beam. The measured volume is coloured following the Poincaré sphere and reveals the topological structure of the Hopf fibration (**a**). Two C lines, the black loop and the threading straight white line, organise the texture into nested tori. Each toroidal surface represents points characterised by the same ellipticity. The colours wind nontrivially around each torus, and a few polarization filaments making up these tori are shown in the insets: in (**b**), the lighter surface ($S_3 = 0.398$) is made of lines characterised by RH elliptic polarization; in (**c**), the L surface ($S_3 = 0$) is made of lines along which the polarization state is linear[22,25]; in (**d**), the darker surface ($S_3 = -0.775$) is made of lines characterised by LH elliptic polarization. In each inset, the cyan and red filaments, corresponding to $\beta = 0, \pi$ are shown to form a Hopf link. Every pair of filaments in the texture link in this way, consistent with the Hopf fibration.

The VFFR measurements reveal the polarization Hopf fibration in the 3D light structure (Fig. 3a). The polarization ellipticity is constant on nested tori, made up of polarization filaments labelled by constant $\beta$ and varying azimuth $\alpha$ (Fig. 3b-d). Our polarimetric resolution identifies these filaments clearly, particularly the linking between pairs of loops. This resolution compares very well with experimentally measured hopfion structures in other systems, such as cold atoms[11,12] and liquid crystals[10]. As predicted (see Supplementary Information), the two linked C lines (vortices in the superposed beams) are the topological skeleton of the hopfion structure, on which the rest of the polarization texture hangs. They are not topologically privileged–all polarization filaments are linked loops–but the C lines form the core filaments for the system of tori, including the L surface of linear polarization. We anticipate C lines to play a similar structural role in other topological 3D polarization textures.

Considering the shaped beams' phase as well as polarization allows a comparison of the measured hopfion structure in real space (Fig. 4a, with phases along the shown filaments in Fig. 4b) with the optical hypersphere (Fig. 4c), parametrised by $\alpha, \beta, \gamma$. This direct comparison gives a volume-to-volume mapping (demonstrated by the grey cubes in Fig. 4a,c). The density of hypersphere volume with respect to real space volume is the topological Skyrme density $\Sigma$, which can be interpreted as a continuous measure of linking[33] of the polarization filaments. Characteristic of 3D skyrmions[2,5,8], the real space integral of $\Sigma$, concentrated around the C line loop, integrates to a value very close to unity, covering the hypersphere of hypersolid angle $2\pi^2$, i.e. a Skyrme number of 1. The Skyrme number is the degree of the mapping from 3D real space to the hypersphere, corresponding to the element of the homotopy group $\Pi_3$. More details of this are provided in the Supplementary Information.

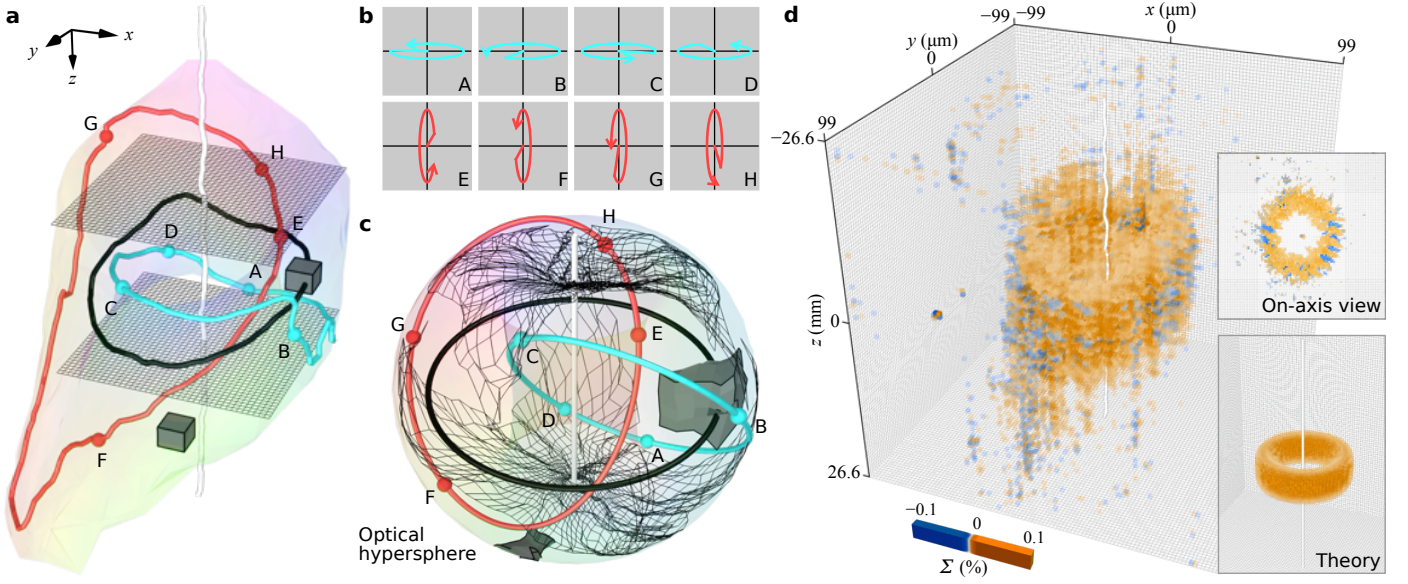

**Figure 4. Measured Skyrme density and optical hypersphere.** The experimental polarization hopfion is shown (**a**), with measured phases (**b**) around the two filaments shown ($\alpha = 0, \pi$, $S_3 = 0.398$, as in Fig. 3b). The hopfion in real space closely resembles the structure of the optical hypersphere parameter space in (**c**), shown in volume-preserving projection from the RH circular polarization state at the focal point of real space. Several features make the nature of the topological mapping clear. The real-space filaments of constant polarization are mapped to the smooth Hopf circles of fixed polarization in the optical hypersphere. The images of two typical real space transverse planes (grey grids) are distorted in the parameter space. The cube in real space intersecting the LH C line (black loop) maps to a larger distorted cuboid, indicating a greater Skyrme density $\Sigma$ near this point. The cube away from the loop maps to a smaller cuboid, indicating a smaller $\Sigma$. In (**d**), the bounding cube represents the investigated focal volume in real space. The positive Skyrme density $\Sigma$ of our structured skyrmionic hopfion is concentrated around the LH C line with some positive and negative fluctuations visible around it. The upper inset shows the on-axis view (from $z = +\infty$) of the toroidal conformation. The measured Skyrme number, given by the sum of the cubes volume, is 0.942 (described in Methods). Theoretical predictions of the Skyrme density for the model field are shown in the lower inset, giving a Skyrme number of 0.997 (calculations in the Supplementary Information).

Mathematically, the Skyrme density $\Sigma$ is the jacobian determinant of the map from real space to the hypersphere (see Supplementary Information),

$$\Sigma = \tfrac{1}{16\pi^2}\nabla\gamma \cdot (\nabla\cos\beta \times \nabla\alpha). \qquad (2)$$

This is the natural 3D generalisation of the 2D topological density for 2D skyrmions[13,15] (here, full Poincare beams), $\tfrac{1}{4\pi}\hat{\boldsymbol{z}} \cdot (\nabla\cos\beta \times \nabla\alpha)$. As the field parameters vary longitudinally as well as transversely, three parameters are needed to determine the full, continuous topological density determining the covering of the optical hypersphere, which is nonzero when the three gradient vectors are linearly independent. The topological density in equation (2) may be rewritten in terms of the normalized optical orbital current[29] $\boldsymbol{J}_{\mathrm{o}} = \mathrm{Im}[E_{\mathrm{R}}^{*}\nabla E_{\mathrm{R}} + E_{\mathrm{L}}^{*}\nabla E_{\mathrm{L}}]$,

$$\Sigma = \tfrac{1}{4\pi^2}\boldsymbol{J}_{\mathrm{o}} \cdot \nabla \times \boldsymbol{J}_{\mathrm{o}}. \qquad (3)$$

Details are given in Supplementary Information. An analogous expression applies to 3D skyrmions in other systems[3,7], with an appropriate current or velocity substituted. It is also the topological helicity, describing knotted fields in high energy physics[2], superfluids[3,7], magnetic fields and hydrodynamics[34]. Its appearance in equation (3) suggests a relation between the 3D Skyrme density of a polarization field and the Poynting vector of optical energy flow.

We determine the Skyrme density explicitly from the measured data, as shown in Fig. 4d. The sum over the measured voxels gives a Skyrme number of 0.945, which is less than unity since low intensities limit the measured volume boundary. The corresponding covering of the optical hypersphere, with the image of the real space measurement boundary, is represented in Supplementary Fig. 10. Rather than a smooth interpolation of the optical field measurements, this density is determined discretely from a simplicial cell complex of spherical tetrahedra in the optical hypersphere arising from the measured data points. Details of the technique and its implementation are in the Methods and Supplementary Information. The value of the Skyrme number of the theoretical field, with the same boundary, is 0.997, consistent with the experimental error.

**Discussion**

We have demonstrated the experimental construction of a 3D skyrmionic hopfion in the polarization and phase pattern of a propagating light beam. The Hopf fibration is realised in the natural polarization parameters from equation (1), a mapping from 3D real space to the 3D optical hypersphere, generalising the Poincaré sphere naturally by including phase.

Our experiment and analysis manifest several topological ideas not commonly emphasised in optics. Firstly, optical polarization fields in 3D can have topological textures, analogous to textures in condensed matter, high-energy physics, etc. This might lead to further insights and possibilities for topologically structured light and its applications. Secondly, as a parameter space for the full vectorial light field, the optical hypersphere goes beyond the standard Poincaré sphere. The usual approach requires a Pancharatnam-Berry phase[35,36] to be included later, ignoring the fact that the optical field parameters define a manifold as natural as the 3-sphere. It is intriguing to speculate whether the machinery of the Poincaré sphere analysis of polarization and Jones calculus may be cast in the optical hypersphere.

The 3D polarization Skyrme density $\Sigma$ in equations (2) and (3) is a new tool to analyse optical vectorial full fields. $\Sigma$ is the continuous topological charge density representing the abstract optical hypersphere volume covered by each real space point. $\Sigma = 0$ when the gradients of ellipticity, phase, and azimuth are linearly dependent, typically occurring along surfaces in 3D. The relation between $\Sigma$ and the optical orbital current suggests a subtle interplay between the Poynting vector[29] and energy-momentum fluxes with optical hypersphere topology (explored further in Supplementary Information).

A smooth polarization texture is disrupted at point singularities in the polarization field, such as saddle points in the parameters $\alpha, \beta, \gamma$. As previously observed[24] in the reconstruction of Seifert surfaces spanning knotted optical singularities, these points are experimentally hard to control and limit the effective reconstruction of textures of polarization lines. They do not affect the Skyrme number, and such points will lie on the surfaces $\Sigma = 0$.

Our experiments and theory demonstrate new topological invariances possible in structured light. This formulation and measurement of an optical $\Pi_3$ invariant will lead to new, robust topological design principles for 3D optical fields for free space optics and nanophotonics. These skyrmionic structures generalise to fields with higher degree Skyrme numbers, involving more complex superposed beams including knots and links, offering a broader gamut of topological structures and integers that can be encoded in structured optical beams. This approach to topological beam shaping will offer further analogies with cold atoms, condensed matter and high energy physics, offering the possibility of emulating, optically, exotic particle-like topologies from field theories not accessible otherwise in the laboratory.

**Acknowledgements:** We are grateful to Miguel Alonso, Michael Berry, Jörg Götte, Michael Morgan, Renzo Ricca, Paul Sutcliffe, Benny Jung-Shen Tai, Alexander Taylor, Teuntje Tijssen, Jonathan Watkins, Alessandro Zannotti and Shuang Zhang, and especially Benjamin Bode, David Foster and Ivan Smalyukh for conversations, advice and research support. The numerical computations were performed using the University of Birmingham's BEAR Cloud service. DS and MRD acknowledge financial support from the University of Birmingham, the Leverhulme Trust Research Programme RP2013-K-009 (SPOCK: Scientific Properties of Complex Knots) and the EPSRC Centre for Doctoral Training in Topological Design (EP/S02297X/1). RR, DE, EO, and CD acknowledge partial support by the German Research Foundation (DFG), under project DE 486/22-1 and DE 486/23-1, as well as by the European Union (EU) Horizon 2020 program, in the framework of the European Training Network ColOpt ITN 721465. JR acknowledges financial support from Engineering and Physical Sciences Research Council (EP/S002952/1 and EP/P026133/1). DS and FN are supported in part by: Nippon Telegraph and Telephone Corporation (NTT) Research, the Japan Science and Technology Agency (JST) [via the Quantum Leap Flagship Program (Q-LEAP), the Moonshot R&D Grant Number JPMJMS2061, and the Centers of Research Excellence in Science and Technology (CREST) Grant No. JPMJCR1676], the Japan Society for the Promotion of Science (JSPS) [via the Grants-in-Aid for Scientific Research (KAKENHI) Grant No. JP20H00134 and the JSPS–RFBR Grant No. JPJSBP120194828], the Army Research Office (ARO) (Grant No. W911NF-18-1-0358), the Asian Office of Aerospace Research and Development (AOARD) (via Grant No. FA2386-20-1-4069), and the Foundational Questions Institute Fund (FQXi) via Grant No. FQXi-IAF19-06.

**Author contributions:** DS and RD equally share first authorship. DS and MRD formulated the theory and developed the numerical methods with assistance from JR; RD and EO designed and performed the experiment, supported by DE; CD, JR and MRD provided explanations of data; FN, CD and MRD supervised the project.

## METHODS

### Topological design of optical skyrmionic hopfion

The optical skyrmionic hopfion consists of the two scalar fields $E_R$ and $E_L$ representing the right- and left-handed field components respectively. These scalar components are appropriately structured to give the 3D topological texture described in the main text, effectively realising the topological mapping from 3D real space to the optical hypersphere. In these methods, we will refer to unnormalized field amplitudes $\psi_R$ and $\psi_L$ rather than their normalized counterparts: the beam intensity is $I = |\psi_R|^2 + |\psi_L|^2$, and $E_j = \psi_j / \sqrt{I}$, $j = R, L$.

The optical skyrmionic hopfion can be understood intuitively quite simply: the component $\psi_R$ should have a circular optical vortex line in the focal plane, concentric to the beam axis, and the component $\psi_L$ should have an optical vortex line along the beam axis. This realises all phases and polarizations (i.e. all points of the optical hypersphere), concentrated in a small propagation volume. These conditions can be realised by superpositions of Laguerre-Gaussian (LG) modes[22,29]. The standard definition of these modes (given in Supplementary Information) defines $LG_{\ell,p}(R, \phi, z; w)$, depending on cylindrical coordinates in real space, $(R, \phi, z)$, with $\ell$ the azimuthal mode number, $p$ the radial mode number, and $w$ the waist width.

As discussed in the Supplementary Information, the axial optical vortex line should have a negative sign, so we choose $\psi_L = 2c\,LG_{-1,0}(R, \phi, z; w)$, the simplest LG mode with an axial vortex of the correct sign, with $c$ a constant to be found and $2$ included for calculational convenience. The vortex ring can be realised by the sum of two LG modes with $\ell = 0$, $\psi_R = (-a + b)LG_{0,0}(R, \phi, z; w) - bLG_{0,1}(R, \phi, z; w)$, where $a$ and $b$ are parameters to be found. This guarantees the vortex ring to be in the focal plane $z = 0$, with a radius of $\sqrt{a/b}\,w$, provided $a, b > 0$. The coefficients $a$ and $b$ determine the intensity pattern around the vortex ring as well as its radius.

For fixed value of $w$ the optical skyrmionic hopfion is therefore realised for a range of values of $a, b$ and $c$. The different values of the parameters give very different shapes of the structure (residing in the polarization parameters) and distribution of the overall intensity $I$. A preliminary exploration of these is given in the Supplementary Information. The spreading nature of the gaussian beams means that it is not possible to cover the 3-sphere completely with polarization states realised in 3-space. We therefore choose a superposition which maximises the volume of optical states in the optical hypersphere within the measured 3D volume in real space.

To be effectively generated and measured in the experiment, the values of the parameters are chosen to optimise the field configuration. To aid this optimisation, we introduce an extra scale size parameter $\mathcal{K}$, with $b = b_0 \mathcal{K}^2$ and $c = c_0 \mathcal{K}$. The 3D size of the skyrmionic hopfion scales according to $\mathcal{K}$, where now the vortex ring radius is $R_0 = \sqrt{a/b}\,(w/\mathcal{K})$. The remaining parameters $a, b_0, c_0$, determine the particle-like field distribution's shape. The parameters $a, b_0, c_0$ and $\mathcal{K}$ were chosen to ensure the experimental skyrmionic hopfion to be localised within the measured volume, in practice a cartesian cuboid centred around the focal point. We optimised against the criteria in the following list. (i) Vortex ring radius $R_0$ not larger than beam waist $w$ (this principle is also used in the design of optical vortex knots[37,38]). (ii) Concentrate intensity inside the measured volume, with $I \approx 0$ outside the measured volume. It was especially important to localise the intensity within the transverse cross-section, so as not to lose critical polarization information. (iii) Distribute the intensity as evenly as possible within the measured volume. To maximise the quality of the measured polarizations we avoided regions of low intensity as much as possible where the polarization state changes rapidly. (iv) Concentrate the Skyrme density (continuous topological charge density) within the measured volume, i.e. $\Sigma \approx 0$ outside the measured volume. The density $\Sigma$ is given in main text equations (2) and (3), and described in detail in the Supplementary Information. This enables a measured value of the Skyrme number very close to 1, as described in the remainder of the Methods.

We proceeded by making an estimate of the parameters based on the topological 3D plots of the numerical models, and then improved these based on the quality of the experimental measurements.

The experimental setup, as described below, requires the Fourier transform of the beam superposition to be realised on the SLM, and the desired field is mathematically back-propagated through the paraxial lens system using Fourier optics[39]. The LG distributions when $z = 0$ are eigenfunctions of the Fourier transform operation. Thus the real space LG mode $\mathrm{LG}_{\ell,p}(R, \phi, z; w)$ corresponds, in Fourier space, to the 2D amplitude $\mathrm{i}^{2p-|\ell|}\mathrm{LG}_{\ell,p}^{2D}(\boldsymbol{q}_\perp; w_\mathcal{F})$, where $w_\mathcal{F}$ is the corresponding waist in the Fourier plane with transverse position $\boldsymbol{q}_\perp$. The holograms correspond to $-2\mathrm{i}c_0\mathcal{K}\mathrm{LG}_{-1,0}^{2D}(\boldsymbol{q}_\perp; w_\mathcal{F})$ for $\psi_\mathrm{L}$ and $(-a + b_0\mathcal{K}^2)\mathrm{LG}_{0,0}^{2D}(\boldsymbol{q}_\perp; w_\mathcal{F}) + b_0\mathcal{K}^2\mathrm{LG}_{0,1}^{2D}(\boldsymbol{q}_\perp; w_\mathcal{F})$. The coefficients do not depend on the Fourier waist $w_\mathcal{F}$, so the overall beam in real space scales linearly in radius $R$ and quadratically in propagation distance $z$ as $w_\mathcal{F}$ is varied. This quantity is chosen so that the skyrmionic hopfion has the desired size in real space whilst fully utilising the SLM.

In our optical system, $\lambda = 532\,\mathrm{nm}$, the waist of the beam on the SLM is $w_\mathcal{F} = 6.252 \times 10^{-4}\,\mathrm{m}$, and the imaging system given by lenses L1 and L2 (Supplementary Fig. 4b) halves the size of the beam. The resulting waist width is $w = 54.2\,\mathrm{\mu m}$ (giving a Rayleigh range $z_\mathrm{R} = 34.7\,\mathrm{mm}$). The measured volume is a cuboid, $|x| \leq x_{\max}$, $|y| \leq y_{\max}$, $|z| \leq z_{\max}$, with $x_{\max} = 3.13w = 170\,\mathrm{\mu m}$; $y_{\max} = 3.91w = 212\,\mathrm{\mu m}$; $z_{\max} = 0.768z_\mathrm{R} = 26.6\,\mathrm{mm}$. The values for the beam parameters were optimised in this range to be $a = 3$, $b_0 = 1.5$, $c_0 = 0.16$, $\mathcal{K} = 2.5$. In terms of the original parameters, this gives the values $a = 3$, $b = 9.4$, $c = 0.4$. With these choices, the LH C line ring is at $R_0 = 0.57w = 30.6\,\mathrm{\mu m}$. The field configuration of this model field near the C line ring is shown in Supplementary Fig. 2b, resembling the corresponding Hopf fibration configuration (e.g. Supplementary Fig. 2a) closely.

## Optical system design

The experimental skyrmionic hopfion field is the superposition of two structured beams of orthogonal circular polarization, $\psi_\mathrm{R}$ and $\psi_\mathrm{L}$. Experimentally, these two scalar components are shaped by the amplitude and phase modulation of a collimated laser beam (horizontal linear polarization, expanded) performed by a reflective phase-only spatial light modulator (SLM; Holoeye Pluto phase-only, $1920 \times 1080$ px HD display), shown in Supplementary Fig. 4b. The SLM is used in split-screen mode[40,41,42], with each half embedding the amplitude and phase information of $\psi_\mathrm{R}$ and $\psi_\mathrm{L}$ respectively. To optimise the beam quality, the Fourier hologram for each polarization component is a $600 \times 600$ pixels square. This resolution was proven to produce all details of the transverse beam structure in the focal volume. The two holograms are placed so each receives approximately homogeneous illumination of the expanded input laser beam without losing too much intensity. The phase-only hologram is shown in Supplementary Fig. 4a.

To allow for amplitude modulation by a pure phase hologram, a weighted blazed grating is applied[43]. The desired scalar modes appear in the first diffraction order, which is spatially filtered by an aperture A in the conjugate plane of the SLM, generated by lens L1 (shown in Supplementary Fig. 4b). Fourier holograms are applied on the SLM, so that the desired beams are sculpted in the focus of the Fourier lens (L1), i.e. in the conjugate plane of the SLM. The hologram for each beam is normalized separately, taking advantage of the full modulation depth of the SLM for each beam individually.

The two beams are subsequently combined on-axis by an interferometric system. Before they are combined, the two beams are given orthogonal linear polarizations by a combination of a half wave plate (HWP) and a polarizing beam splitter (PBS), allowing also for the adjustment of the beams' intensity ratio. This is a critical step to realise the complex polarization structure: the HWP angle directly affects the relative strength of the two components and hence the coefficient $c$ in the field design described above.

After the beams are combined, a quarter wave plate (QWP) transforms the orthogonal linear polarization states into orthogonal circular polarizations. The imaging system given by lens L1 and L2 (Supplementary Fig. 4b) halves the size of the beam and L3 performs the final Fourier transform that gives the skyrmionic hopfion in its focal volume. The focal structure is magnified by lens L4 (16×) onto a CMOS camera (Cam; uEye SE (UI-1240SE), 1280 × 1024 px).

**Volumetric full field reconstruction**

We retrieve the full field information (transverse components of the paraxial beam) by reconstructing the polarization and phase in the focal volume. Supplementary Fig. 5 shows five transverse planes at different positions in the propagation direction for the normalized Stokes parameters $S_1$, $S_2$, $S_3$ and the phases $\chi_R$ and $\chi_L$ of the RH and LH field components. The measurements in multiple transverse planes are performed via digital propagation[32] (see Supplementary Information). A detailed description of polarimetry[44] (Supplementary Fig. 4c) and transverse phase interferometry[45] (Supplementary Fig. 4d) can be found in the Supplementary Information. The polarization measurements across different planes are unaffected by the harmonic time dependence of the optical field and are directly stored into 3D arrays. However, when stacking volumetric phase measurements, the relative phase between neighbouring planes must be retrieved. First, we describe our procedure for connecting the transverse phase measurements to their neighbouring planes, and then we present our routine to minimise the experimental error in retrieving the field components.

The measured transverse phase structure per plane is constituted of the light field's propagation term, $e^{ikz}$ times the superposed LG structure described in the "Topological design" Method section above. This includes a Gouy phase factor $e^{-it\chi^G}$, where $\chi^G(z/z_R)$ is the $z$-dependent Gouy Phase, and a phase term varying radially and longitudinally (full Laguerre-Gauss modes equation is given in Supplementary Information), and a time-dependent phase offset due to the time varying phase relation between the measured and reference beams. In order to concentrate on the transverse variation, we circumvent the effect of $e^{ikz}$ within the measurements per $z$-plane, thereby avoiding the effects of undersampling the electric field oscillation, by setting the distance between two transverse planes to a multiple of the wavelength ($100\lambda$), so the propagation factor $e^{ikz}$ is negligible. Next, we choose a transverse reference point ($\boldsymbol{r}_{\perp \mathrm{ref}}, z$) close to the optical axis ($R \approx 0$), so that the phase at this point is only affected by the $z$-dependent Gouy phase term of the LG beams and is unaffected by the other spatially varying phase factors. For each plane, the phase of the reference point is set to the same value, so the Gouy phase and the time-dependent phase offset are subtracted. In order to finalise the missing relation between different $z$-planes, the theoretical Gouy phase term $\chi^G$ is added. Note that the Gouy phase represents an offset value per plane, only depending on the $z$-position but without any dependence on the transverse coordinates. Thus, the measurements themselves are not affected by this approach and, as a result, we correct for the errors in $z$ caused by the time-dependent variations in the measurement system. Supplementary Fig. 6 shows the $x = 0$ plane (longitudinal cut) of the theoretically expected (left) and the reconstructed (right) 3D phase structures of $\chi_R$ and $\chi_L$. This figure demonstrates that the reconstructed 3D phase distributions are consistent with the theoretical predictions.

Due to experimental errors (see Supplementary information), the singularities of the differences of the phase of the two field components $\chi_L - \chi_R$ (wrapped between $-\pi$ and $\pi$) do not coincide with those of the polarimetrically-determined $\arctan(S_1, S_2)$ as the polarization and phase measurement are independent. Observations of the 3D structure of the C lines from the polarization measurements and the phase singularities from the phase measurements allow the systematic error to be minimised by shifting the polarization measurements until the C line loop coincides with the singular loop of $\chi_R$. Moreover, the overall error is reduced by redefining the Stokes parameters $S_1$ and $S_2$ as follows: $S_1 = 2\sqrt{s_0^2 - s_3^2} \cos(\chi_L - \chi_R)/s_0$ and $S_2 = 2\sqrt{s_0^2 - s_3^2} \sin(\chi_L - \chi_R)/s_0$. To finalise the volumetric full field reconstruction we calculate the real and imaginary parts of the beam components from $E_R = \sqrt{(s_0 + s_3)/2}\, e^{i\chi_R}$, and $E_L = \sqrt{(s_0 - s_3)/2}\, e^{i\chi_L}$. The full field is used to calculate the Skyrme density of the optical field as described in the next Method.

**Numerical calculation of experimental Skyrme number**

We measure the Skyrme number of the optical skyrmionic hopfion directly from the discretely sampled, measured data by taking advantage of the robustness of topology. This optimises the computational speed necessary to evaluate the Skyrme number from experimental measurements. The measured polarization and phase at each point in real 3D space correspond to a point in the optical hypersphere. The 3D cubic lattice of measured voxels is mapped into a topology-preserving but distorted lattice in the optical hypersphere. An example for the ideal hopfion field (see Supplementary Information), is shown in Supplementary Fig. 9a-b. The measured Skyrme number is therefore based on this piecewise-linear mapping generated from the measured data points without interpolation. This approach can readily be used for measurements of other physical Skyrme-like maps, including lower dimensional ones (e.g. via triangular meshes).

The fully resolved experimental data in the focal volume gives real space voxels forming a cuboidal grid. We are interested in the Skyrme density of the real space volume given by a cuboid with transverse size $L_\perp = 1.84w = 99.4\ \mu m$ and longitudinal size $L_\parallel = 0.768z_R = 26.6\ mm$ as defined in the Supplementary Information. Since the image of the cuboidal mesh covers the volume of the hypersphere, reducing the resolution maintains this filling. The numerical routine is made more time efficient by reducing the resolution to a cubic mesh of dimension $101 \times 101 \times 101$ in physical space, centred at the focal point. The voxels are centred at points labelled by $(i, j, k)$ with $1 \le i, j, k \le 100$. Each such point corresponds to a normalized 4D vector $\vec{n} = (\text{Re}E_R, \text{Im}E_R, \text{Re}E_L, \text{Im}E_L)$ found via the VFFR method, giving a distorted cubic 3D grid in the optical hypersphere whose vertices are the points $\vec{n}(i,j,k)$. The distortion of the experimentally measured field is significantly greater than the example in Supplementary Fig. 9a-b, as can be seen in the images of the two real space planes in the optical hypersphere in main text Fig. 4. Each elementary cube $C = C_{i,j,k}$ is labelled by $i,j,k$, with vertices $\vec{n}(i,j,k)$, $\vec{n}(i+1,j,k)$, $\vec{n}(i,j+1,k)$, ... denoted by $c1$, ..., $c8$ as indicated in Supplementary Fig. 9c-d. The cube $C$ occupies a volume $\mathrm{Vol}(C)$ within the optical hypersphere. We numerically determine $\mathrm{Vol}(C)$ as follows.

An elementary topological cell in 3D is a tetrahedron (i.e. a 3-simplex[46], in the language of simplicial topology). We convert our cubic $i,j,k$ lattice into a 3D simplicial complex by decomposing each cube into five irregular tetrahedra. The resulting mesh of tetrahedra, where neighbours share triangular faces, edges and vertices, make up a 3D cell complex[46]. The tetrahedra can share the cube's vertices in two distinct ways, which are given by the following ordered sets of four vertices (see Supplementary Fig. 9c-d): (A), ($c1$, $c2$, $c4$, $c5$), ($c4$, $c5$, $c7$, $c8$), ($c5$, $c2$, $c7$, $c6$), ($c2$, $c7$, $c3$, $c4$), ($c5$, $c7$, $c2$, $c4$) and (B), ($c1$, $c2$, $c3$, $c6$), ($c3$, $c4$, $c1$, $c8$), ($c5$, $c6$, $c1$, $c8$), ($c7$, $c8$, $c3$, $c6$), ($c8$, $c1$, $c3$, $c6$). For any cubic lattice, cubes can be decomposed into two choices of tetrahedral mesh: cubes of type A at positions where the quantity $i+j+k$ is even (odd) and cubes of type B where $i+j+k$ is odd (even). We compute both types of 3D cell complex as a check of numerical accuracy. As a result, the measurement points in real space and measured values in the hypersphere define a piecewise linear map representing the physical field.

In the hypersphere, the tetrahedra are constructed so the edges joining the vertices are geodesics. Each tetrahedron's four faces are spherical triangles, and along edges, pairs of faces meet at the dihedral angles $0 < \varphi_j < \pi$, for $j = 1, 2, ..., 6$. The formula for the 3D volume $\mathrm{Vol}(T)$ of an irregular spherical tetrahedron $T$ constructed in this way can be written explicitly in terms of dihedral angles by means of Murakami's formula[47] (see Supplementary Information). The contribution to the Skyrme number comes from the signed volumes $\mathrm{sign}[\det(\vec{n}_a, \vec{n}_b, \vec{n}_c, \vec{n}_d)]\mathrm{Vol}(T)$, where $\vec{n}_\ell$ with $\ell = \{a, b, c, d\}$ are 4D unit vectors pointing to the four vertices of a spherical tetrahedron $T$. Only tetrahedral cells included within a 3-dimensional hemisphere, whose volume is less than $\pi^2$, are considered. The sign of the volume comes from the ordering of the vertices with respect to the right-hand rule, where the triangular base $a, b, c$ follows the fingers and the vertex $d$ follows the thumb. When the volume is negative, the order of the vertices in real space and that of the vertices of the tetrahedron in the optical hypersphere are inverted. This follows the standard orientation rules of a 3-simplex.

At higher resolutions of the cubic lattice, the spherical tetrahedra are smaller, and the curved edges tend to become linear and the spherical distortion can be neglected: the tetrahedron volume are better approximated by its flat-space analogue. The 101×101×101 mesh defined above is consistent with the volume of the tetrahedra being within the range allowed by numerical precision.

The hyperspherical volume of the cube $C$ corresponds to the union of the volumes of the associated, neighbouring tetrahedra comprising $C$, and its volume $\mathrm{Vol}(C)$ is the sum of the signed spherical tetrahedra volumes $\mathrm{Vol}(T)$. The results for each such $\mathrm{Vol}(C_{i,j,k})$ are stored in two 3D arrays, one for each type of 3D cell complex. The experimental Skyrme number is found by adding together the volumes of all the hypersphere cubes with appropriate sign, normalized by the 3-sphere volume: $\sum_{i,j,k} \mathrm{Vol}(C_{i,j,k})/(2\pi^2)$ (see Supplementary Information). The measured 3D Skyrme number corresponds to the fraction of the hypersphere volume by the image of the measured volume of real space. The sums over all the elements in the arrays give

Skyrme numbers 0.94521 and 0.94528 for the two kinds of mesh. We take the experimental Skyrme number to be the mean of these two numbers, 0.94524. It is straightforward to implement the vector calculation described here in a numerical algorithm in MATLAB or Python. The volumes of the cubes in the meshes can be calculated in parallel via high performance computers.

## REFERENCES

1. Skyrme, T. H. R. A non-linear field theory. *Proc. R. Soc. A* **260**, 127-138 (1961).
2. Manton, N. & Sutcliffe, P. *Topological solitons. Cambridge University Press* (2004).
3. Volovik, G. E. & Mineev, V. P. Particle-like solitons in superfluid $^3$He phases. *JETP Lett.* **46**, 401-404 (1977).
4. Weeks, J. *The shape of space*. Marcel Dekker, New York (1985).
5. Faddeev, L. D. & Niemi, A. J. Knots and particles. *Nature* **387**, 58-61 (1997).
6. Cruz, M., Turok, N., Vielva, P., Martinez-Gonzalez, E. & Hobson, M. A cosmic microwave background feature consistent with a cosmic texture. *Science* **318**, 1612-1614 (2007).
7. Radu, E. & Volkov, M. S. Stationary ring solitons in field theory — knots and vortons. *Phys. Rep.* **468**, 101-151 (2008).
8. Ruostekoski, J. & Anglin, J. R. Creating vortex rings and three-dimensional skyrmions in Bose-Einstein condensates. *Phys. Rev. Lett.* **86**, 3934-3937 (2001)
9. Babaev, E., Faddeev, L. D. & Niemi, A. J. Hidden symmetry and knot solitons in a charged two-condensate Bose system, *Phys. Rev. B* **65**, 100512 (2002).
10. Ackerman, P. J. & Smalyukh, I. I. Diversity of knot solitons in liquid crystals manifested by linking of preimages in torons and hopfions. *Phys. Rev. X* **7**, 011006 (2017).
11. Hall, D. S., Ray, M. W., Tiurev, K., Ruokokoski, E., Gheorghe, A. H. & Möttönen, M. Tying quantum knots. *Nat. Phys.* **12**, 478-483 (2016).
12. Lee, W., Gheorghe, A. H., Tiurev, K., Ollikainen, T., Möttönen, M. & Hall, D. S. Synthetic electromagnetic knot in a three-dimensional skyrmion. *Sci. Adv.* **4**, eaao3820 (2018).
13. Salomaa, M. M. & Volovik, G. E. Quantized vortices in superfluid $^3$He. *Rev. Mod. Phys.* **59**, 533 (1987).
14. Leslie, L. S., Hansen, A., Wright, K. C., Deutsch, B. M. & Bigelow, N. P. Creation and detection of skyrmions in a Bose-Einstein condensate. *Phys. Rev. Lett.* **103** 250401 (2009).
15. Nagaosa, N. & Tokura, Y. Topological properties and dynamics of magnetic skyrmions. *Nat. Nano.* **8** 899–911 (2013).
16. Tsesses, S., Ostrovsky, E., Cohen, K., Gjonaj, B., Lindner, H. & Bartal, G. Optical skyrmion lattice in evanescent electromagnetic fields. *Science* **361** 993-996 (2018).
17. Du, L., Yang, A., Zayats, A. V. & Yuan, X. Deep-subwavelength features of photonic skyrmions in a confined electromagnetic field with orbital angular momentum. *Nat. Phys.* **15** 650-654 (2019).
18. Davis, T. J., Janoschka, D., Dreher, P., Frank, B., Meyer zu Heringdorf, F.-J. & Giessen, H. Ultrafast vector imaging of plasmonic skyrmion dynamics with deep subwavelength resolution. *Science* **386**, eaba6415 (2020).
19. Urbantke, H. The Hopf fibration—seven times in physics. *J. Geom. Phys.* **46**, 125–150 (2003).
20. Thompson, W. (Lord Kelvin) On vortex atoms. *Proc. R. Soc. Edin.* **6**, 94-105 (1867).
21. Beckley, A. M., Brown, T. G. & Alonso, M. A. Full Poincaré beams. *Opt. Exp.* **18**, 10777–10785 (2010).
22. Donati, S., Dominici, L., Dagvadorj, G., Ballarini, D., De Giorgi, M., Bramati, A., Gigli, G., Rubo, Y. G., Szymanska, M. H. & Sanvitto, D. Twist of generalized skyrmions and spin vortices in a polariton superfluid, *PNAS* **113**, 14926–14931 (2016)
23. Dennis, M. R., O'Holleran, K. & Padgett, M. J. Singular optics: optical vortices and polarization singularities. *Prog. Opt.* **53**, 293-363 (2009).
24. Larocque, H., Sugic, D., Mortimer, D., Taylor, A. J., Fickler, R., Boyd, R. W., Dennis M. R. & Karimi, E. Reconstructing the topology of optical polarization knots. *Nat. Phys.* **14**, 1079–1082 (2018).

25. Bauer, T., Banzer, P., Karimi, E., Orlov, S., Rubano, A., Marrucci, L., Santamato, E., Boyd, R. W. & Leuchs, G. Observation of optical polarization Möbius strips. *Science* **347**, 964-966 (2015).

26. Nye, J. F. *Natural focusing and fine structure of light*. IoP Publishing (1999).

27. Wang, J., Yang, J.-Y., Fazal, I. M., Ahmed, N., Yan, Y., Huang, H., Ren, Y., Yue, Y., Dolinar, S., Tur, M. & Willner, A. E. Terabit free-space data transmission employing orbital angular momentum multiplexing. *Nat. Photon.* **6**, 488-496 (2012).

28. Gahagan, K. T. & Swartzlander, G. A. Optical vortex trapping of particles. *Opt. Lett.* **21**, 827-829 (1996).

29. Berry, M. V. Optical currents. *J. Opt.* **11**, 094001 (2009).

30. Yao, A. M. & Padgett, M. J. Orbital angular momentum: origins, behavior and applications. *Adv. Opt. Photon.* **3**, 161-204 (2011).

31. Dennis, M. R. Polarization singularities in paraxial vector fields: morphology and statistics. *Opt. Commun.* **213**, 201-221 (2002).

32. Otte, E., Rosales-Guzmán, C., Ndagano, B., Denz, C. & Forbes, A. Entanglement beating in free space through spin-orbit coupling. *Light: Sci. App.* **7**, 18007 (2018).

33. Whitehead, J. H. C. An expression of Hopf's invariant as an integral. *Proc. Nat. Acad. Sci.* **33**, 117–123 (1947).

34. Moffatt, H. K. Helicity and singular structures in fluid dynamics. *PNAS* **111**, 3663-3670 (2014).

35. Berry, M. V. The adiabatic phase and Pancharatnam's phase for polarized light. *J. Mod. Opt.* **34**, 1401-1407 (1987).

36. Bliokh, K. Y., Alonso, M. A., Dennis, M. R., Geometric phases in 2D and 3D polarized fields: geometrical, dynamical, and topological aspects, *Rep. Prog. Phys.* **82**, 122401 (2019).

37. Dennis, M. R., King, R. P., Jack, B., O'Holleran, K. & Padgett, M. J. Isolated optical vortex knots, *Nat. Phys.* **6**, 118–121 (2010).

38. Sugic D. & Dennis, M. R. Singular knot bundle in light, *J. Opt. Soc. Am. A* **35**, 1987–1999 (2018).

39. Goodman, J. W. Introduction to Fourier Optics, 3rd ed. (Roberts & Company, 2005).

40. Otte, E., Schlickriede, C., Alpmann, C. & Denz, C. Complex light fields enter a new dimension: holographic modulation of polarization in addition to amplitude and phase, *Proc. SPIE* **9379**, 937908 (2015).

41. Alpmann, C., Schlickriede, C., Otte, E. & Denz, C. Dynamic modulation of Poincaré beams, *Sci. Rep.* **7** 10.1117/12.2078724 (2017).

42. Preece, D., Keen, S., Botvinick, E., Bowman, R., Padgett, M. & Leach, J. Independent polarisation control of multiple opticaltraps, *Opt. Exp.* **16**, 15897–15902 (2008).

43. Davis, J., Cottrell, D., Campos, J., Yzuel, M. Y. & Moreno, I., Encoding amplitude information onto phase-only filters. *Appl. Opt.* **38**, 5004-5013 (1999).

44. Schaefer, B., Collett, E., Smyth, R., Barrett, D. & Fraher, B. Measuring the Stokes polarization parameters. *Am. J. Phys.* **75**, 163-168 (2007).

45. Takeda, M., Ina, H. & Kobayashi, S. Fourier-transform method of fringe-pattern analysis for computer-based topography and interferometry. *J. Opt. Soc. Am.* **72**, 156-160 (1982).

46. Hatcher, A. Algebraic Topology (Cambridge University Press, 2002).

47. Murakami, J. Volume formulas for a spherical tetrahedron. *Proc. Am. Math. Soc.* **140**, 3289-3295 (2012).

# Supplementary Information: Particle-like topologies in light

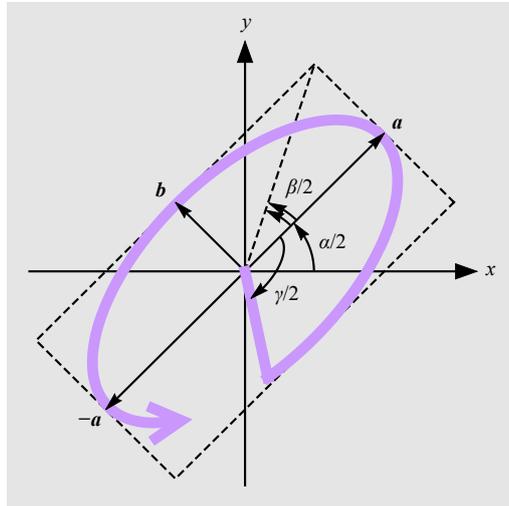

FIG. S1. **Polarization ellipse.** A typical polarization ellipse, bounded by a rectangle. The ellipse has a major axis given by vector $\boldsymbol{a}$ and a minor axis vector $\boldsymbol{b}$; their ordering is consistent the right-handed ellipse orientation (i.e. by $\mathrm{sign}(\boldsymbol{a} \times \boldsymbol{b}) \cdot \hat{\boldsymbol{z}}$). $\boldsymbol{a}$ is chosen rather than its opposite $-\boldsymbol{a}$. The major axis of the ellipse therefore has length $a$ and the minor axis has length $b$, so the bounding rectangle has side lengths $2a$ and $2b$. The normalization $1 = |E_{\mathsf{R}}|^2 + |E_{\mathsf{L}}|^2 = a^2 + b^2$ implies that the bounding box for all ellipses has a fixed diagonal length of $2$. The orientation of the major axis of the ellipse (and hence the longer edge of the rectangle) is $\alpha/2$, defined over a range $-\pi/2 < \alpha/2 \leq \pi/2$. Signed by ellipse handedness, half the area of the bounding rectangle is $\pm 2ab = \cos\beta = S_3$. The phase angle on the ellipse is determined by $\gamma/2$; geometrically, this corresponds to the ellipse auxiliary angle. To determine the full range of phase, $-\pi < \gamma/2 \leq \pi$, it is necessary to pick a reference major axis direction $\pm \boldsymbol{a}$ at which $\gamma = 0$.

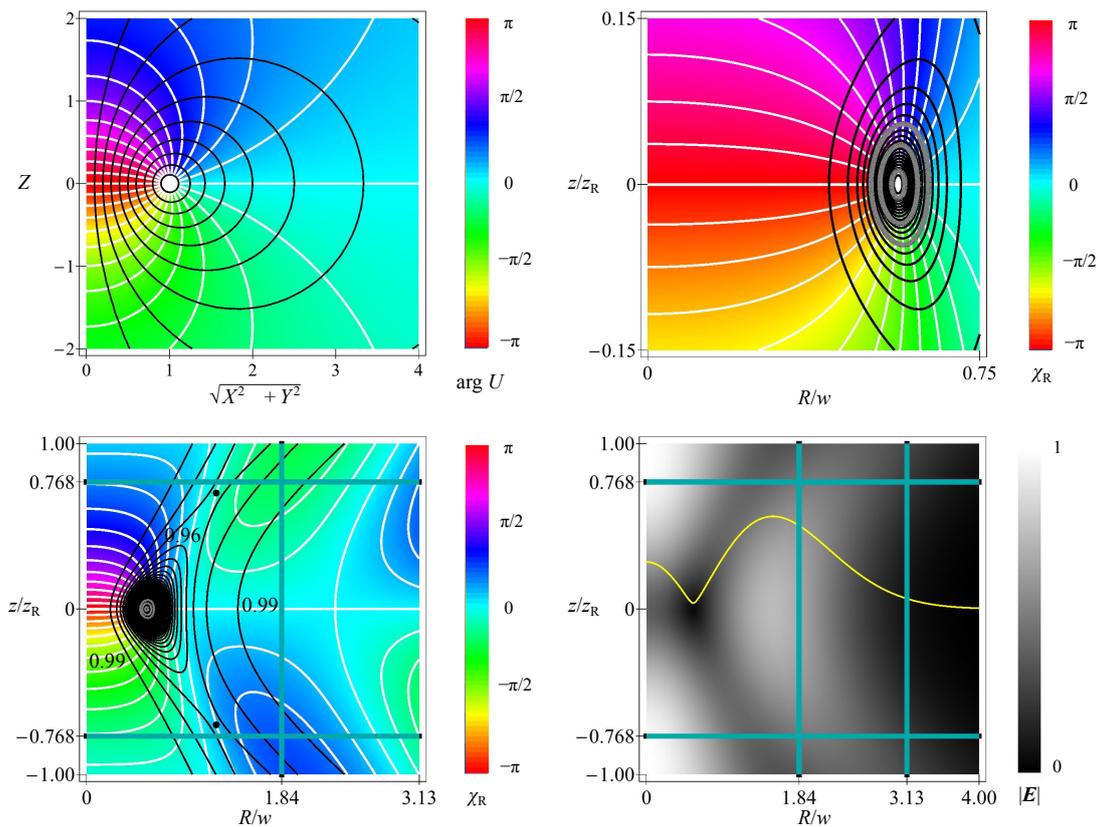

FIG. S2. **Hopfion-like field sections of constant azimuth.** **a**, Section of constant $\varphi$ of the complex field $U(X, Y, Z)$, representing the 3-sphere stereographically projected to infinite euclidean 3-space. The cylindrical radius $\sqrt{X^2 + Y^2}$ is on the horizontal axis, and $Z$ is vertical. The black contours denote contours of $|U|$, and the white lines (and colors) contours of phase $\arg U$. The phase is singular on the nodal line $U = 0$, on which $X^2 + Y^2 = 1$ and $Z = 0$. This is the reference circle of toroidal coordinates $\varphi, \sigma, \tau$, and in fact $\sigma = \arg U$, $\cosh \tau = 1/|U|$, so white and black lines represent toroidal coordinates in this plane. **b**, Section of constant $\phi$ of the model optical field in real space defined in the Methods. The LH C line crosses the plane at its radius $R_0 = 0.57w$. White contours and colors correspond to the axisymmetric phase $\chi_R$ ($\arg E_R$), and black contours correspond to the axisymmetric values of $S_3 = \cos \beta$. The values of $S_3 = -0.775, 0, 0.398$ chosen in the main text are shown in grey, otherwise the contours of $S_3$ are at $-1, -0.9, \ldots 0.9, 1$, increasing from the minimum at the C point. The field strongly resembles the Hopf fibration/toroidal configuration of **a**. **c**, Model optical field with larger $R$ and $z$ range plotted. The contours of $S_3$ are $-1, -0.99, \ldots, 0.99, 1$. The investigated range is indicated by the dark cyan lines showing that the skyrmionic hopfion is mainly inside this range. The Hopf fibration structure breaks down at the saddle points of $S_3$ which are shown as black dots. **d**, Amplitude of the model optical field across the measured range (see Methods). The yellow line shows the real amplitude profile at the focal plane ($z = 0$).

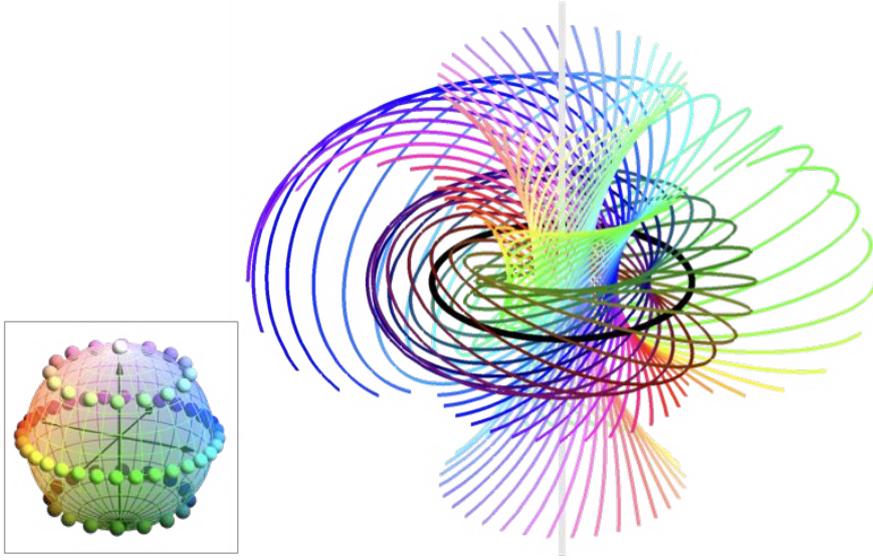

FIG. S3. **Hopf fibration.** The 3-sphere, stereographically projected to 3D euclidean space, is filled by circles. The circles are arranged on tori of constant $\beta$ (equivalently constant $S_3$, constant $\tau$) which are the toroidal surfaces in toroidal coordinates. Each torus corresponds to a fixed height on the fibration's base space – the Poincaré sphere (inset), with the reference circle (black) corresponding the the south pole (LH circular), and the axial line (white) to the north pole (RH circular). Thus each circle corresponds to a point on the 2-sphere, defining the Hopf fibration.

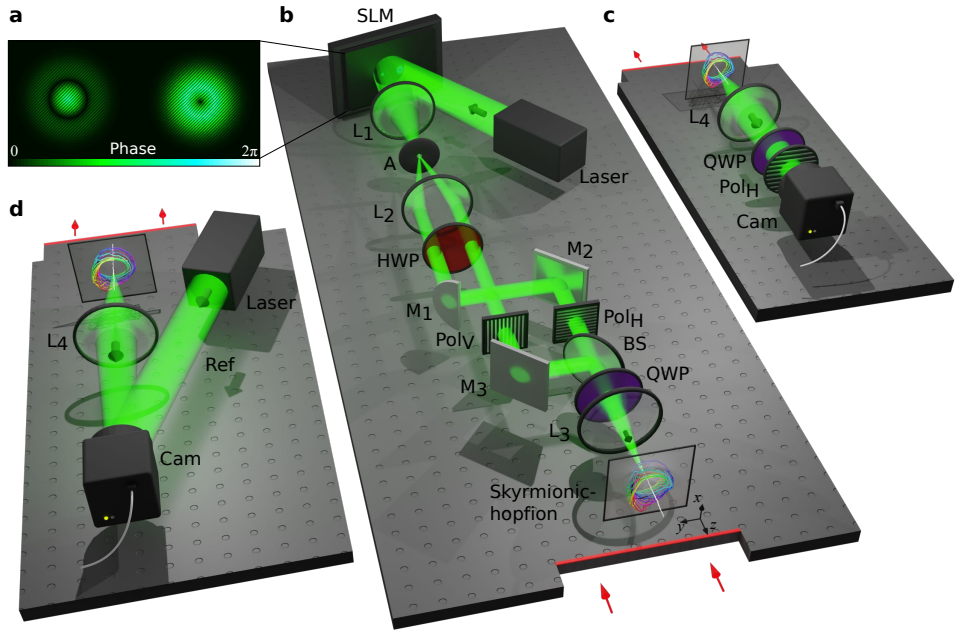

FIG. S4. **Experimental System based on a spatial light modulator (SLM). a**, The amplitude and phase information is encoded as a phase-only Fourier hologram on the reflective SLM. **b**, A weighted blazed grating transfers it to the first diffraction order of the initial laser beam (expanded and collimated; horizontal linear polarization; wavelength $\lambda = 532$nm) when it passes the SLM. The first diffraction order is filtered by an aperture A within a 4f-imaging system (lenses L1 and L2). On-axis superposition of the two beams and their orthogonal circular polarization is ensured by an interferometric system (mirrors M1, M2, M3, beam splitter BS) and polarization optics (H/QWP: half/ quarter-wave plate; PolV,H: vertical/horizontal polarizer). A final lens L3 gives access to the far-field, where the skyrmionic hopfion texture is realised. The VFFR of this structure is based on, **c**, Stokes polarimetry and, **d**, interferometry-based phase measurement (Ref: reference beam) with the corresponding data being collected, transversely resolved, by imaging (lens L4) on a camera (Cam). Longitudinal resolution is achieved by digital propagation enabled by the SLM in combination with the Fourier lens L3. Details on realisation are given in Supplementary Informationand and VFFR is in Methods.

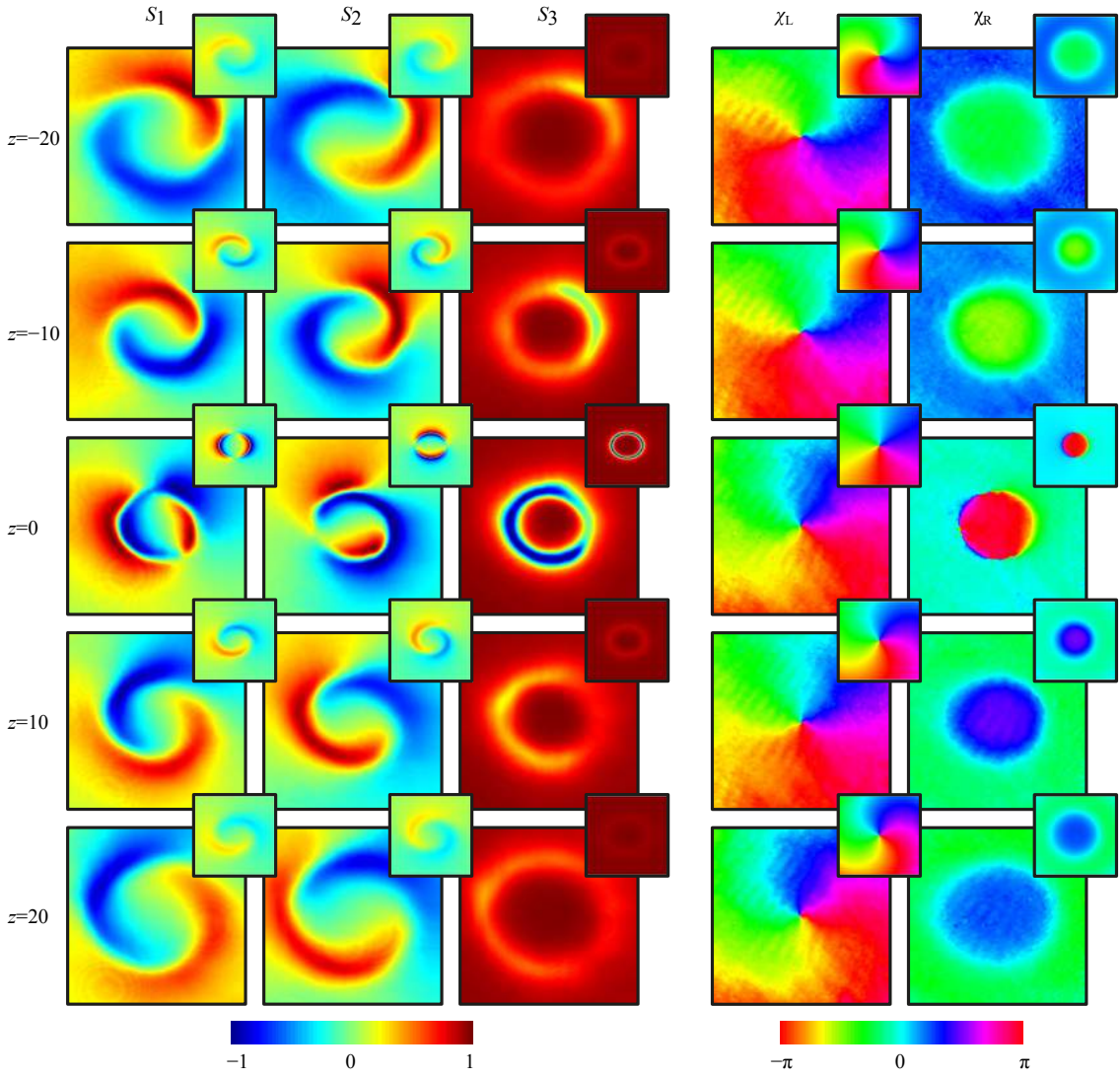

FIG. S5. **Experimental measurements.** Transverse planes along the propagation direction ($+z$) of the normalized Stokes parameters $S_1$, $S_2$, $S_3$ and the phase $\chi_L$ and $\chi_R$. Within the requirements dictated by topology, the measured beam is consistent with the theoretical predictions shown in the insets. The measured planes show $z$ in mm units.

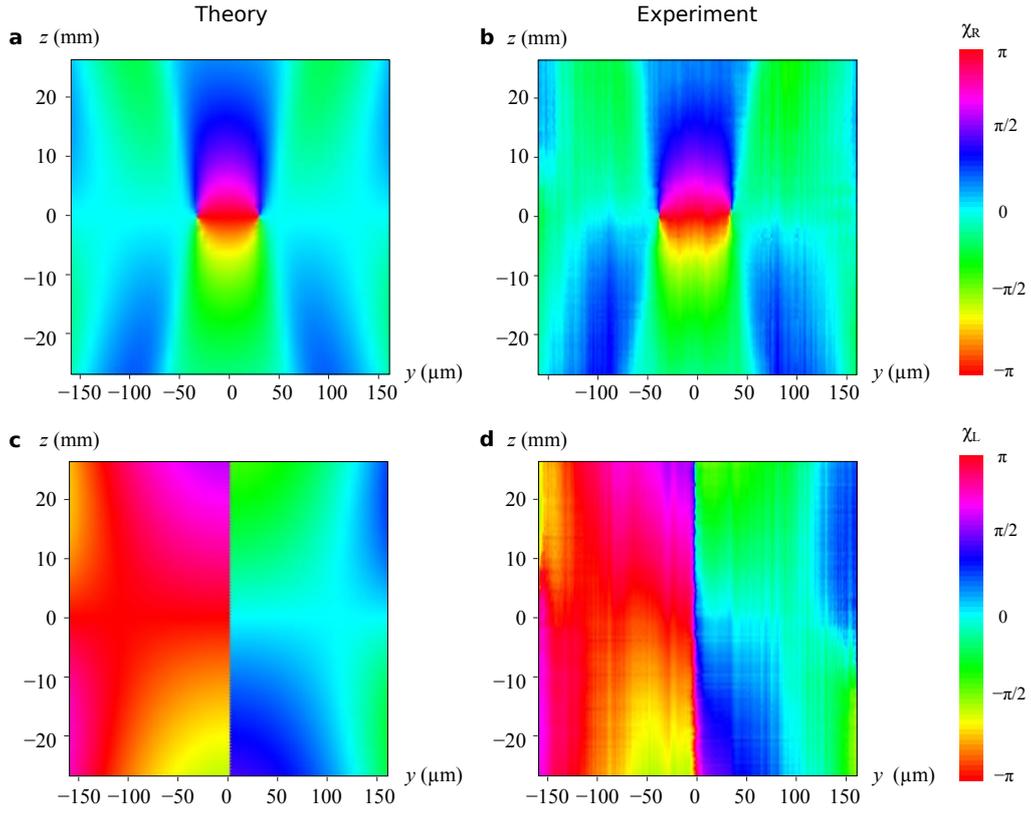

FIG. S6. **Comparison of simulated and experimental phase distributions. a**, **b**, $\chi_R = \arg E_R$; **c**, **d**, $\chi_L = \arg E_L$. The displayed plane is longitudinal, with $x = 0$.

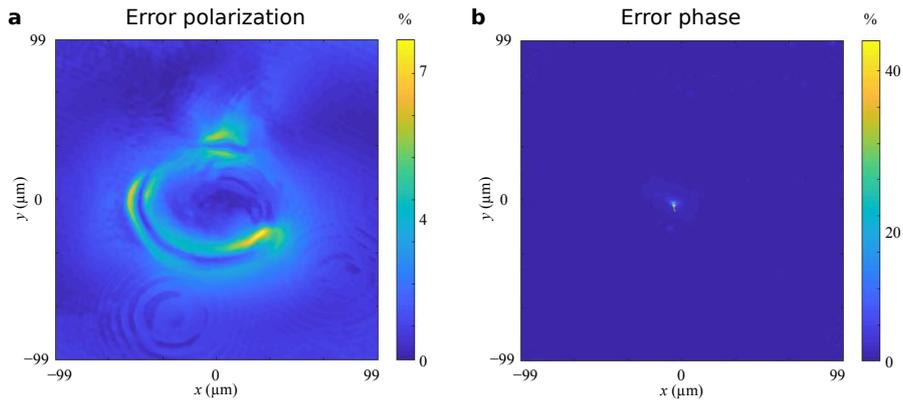

FIG. S7. **Density plot of standard deviation a**, for polarization component $S_3$ and **b**, for LH phase $\chi_L$ at the transverse plane $z = 0$. The basis for this errors are ten measurements of the named parameter in the focal plane. A spatial variation is visible, where the error for $S_3$ is more confined around the nodal line of $E_R$ and the error for $\chi_L$ around the singularity on the optical axis.

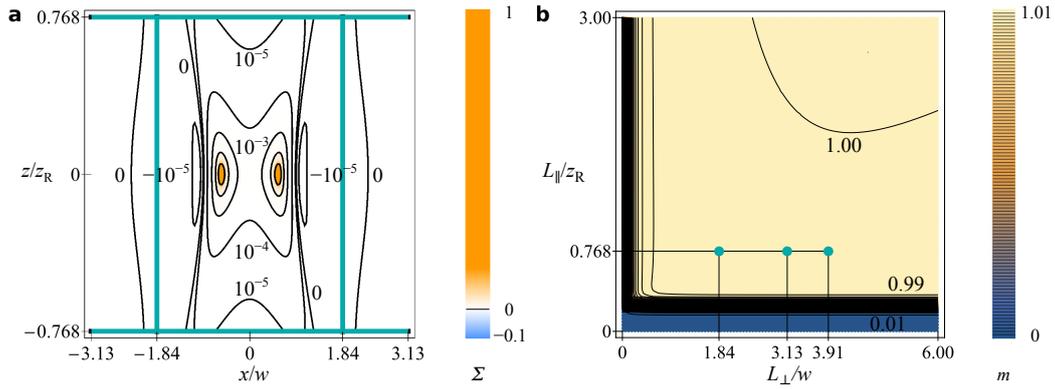

FIG. S8. **Skyrme density and its integral. a**, Skyrme density of the model of the experimental field across the $x, z$ plane. Most of the density is concentrated around the LH C line ring (as shown equivalently in Supplementary Fig. 10b inset), but contours of very low density reveal other structure. The cyan lines indicate the limits of the evaluation of the Skyrme density of the experimental field. Scale bar is with respect to the maximum density $\sim 6 \times 10^{11}\,\mathrm{m}^{-3}$. **b**, Integral of Skyrme density in a cube of size $|x| < L_\perp$, $|y| < L_\perp$ and $|z| \leq L_\parallel$. Provided the integration cube includes most of the orange structure as shown in (a), i.e. $L_\parallel \gtrsim 0.1 z_\mathrm{R}$, $L_\perp \gtrsim R_0$, then most of the integration volume sizes give 0.99, indicative of the particle-like nature of the configuration.

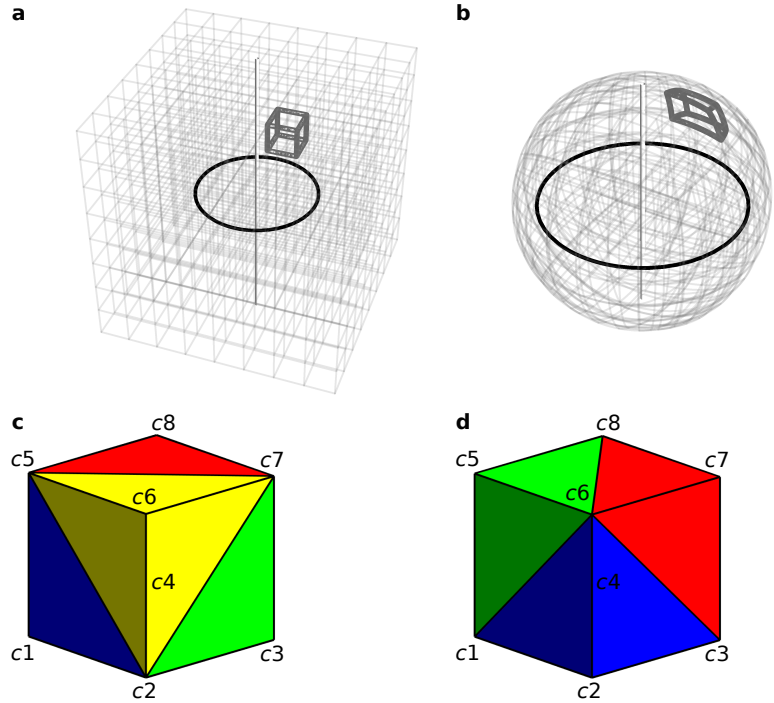

FIG. S9. **Mapping of discrete cubic lattice into the 3-sphere. a**, Real space C line skeleton of ideal skyrmion field $(u, v)$ of Equation (S2.7) overlayed by a cubic lattice in $x, y, z$, whose vertices represent measurement points, from $-2$ to $2$ at each 0.5. The experimentally measured lattice has much greater resolution. **b**, Image of cubic lattice in the optical hypersphere with $u = E_\mathrm{R}$, $v = E_\mathrm{L}$. The hypersphere is represented in volume-preserving projection with center corresponding to $u = -1$, $v = 0$. This is the same projection used in main text Fig. 4b. **c**, **b**, A cubic cell is divided into five tetrahedra with the same vertices: four tetrahedra at corners of the cube, and an internal tetrahedron whose edges are diagonals of the cube faces. The two (mirror image) ways this can be done are shown. For the cubic lattice, alternating between **a** and **b** guarantees neighbouring tetrahedra always share faces (i.e. giving a 3D cell complex).

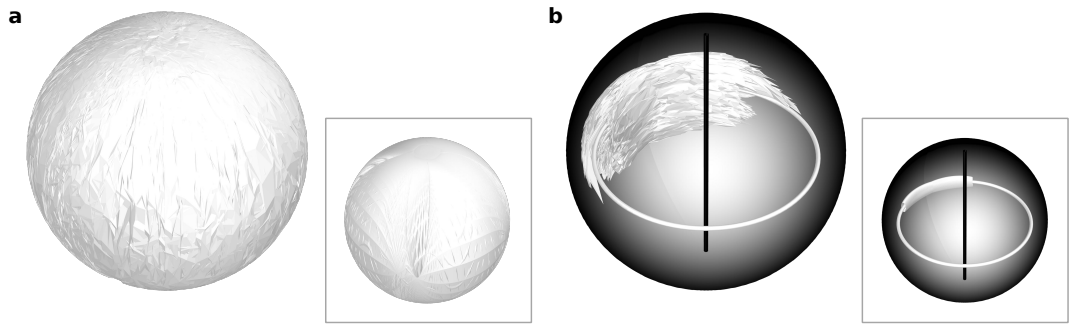

FIG. S10. **The experimental Skyrme number represented as a volume in the optical hypersphere.** The optical hypersphere in these plots is represented by a ball with volume-preserving projection, projecting from different points in **a** and **b**. In each case, the boundary of the measured volume in real space (cube) is mapped to the optical hypersphere as the white surface. The inset shows the corresponding surface for the theoretical model field. The volume bounded by this surface, 94.5% of the total volume, corresponding to the hypersphere volume that is covered experimentally, represents the Skyrme number of 0.945. Insets show the theoretical counterpart, covering 99.7% of the 3-sphere volume. **a,** The central point being the polarization state at the focal point. Most of the ball is filled, with the white surface extending almost to the outer surface. **b,** The projection is now from a LH polarization state. The real space axial C line maps to the white loop, and the white surface encloses the complement of the filled volume.

## Supplementary Note 1: Parametrising the Optical Hypersphere

In this note we discuss the parameters describing the optical state. Starting with the Stokes parameters and Poincaré sphere for polarization, we generalise by including phase to the 3D parameter space we call the *optical hypersphere*. This is the unit 3-sphere in 4-dimensional space. Its natural parametrisation by the polarization and phase parameters defines the Hopf fibration structure. The Poincaré sphere is recovered from the optical hypersphere as the 2-sphere base space of the Hopf fibration.

The state of light at each point in a transverse, monochromatic light beam, travelling in the $z$-direction, is specified by

$$\boldsymbol{E} = E_R \widehat{\boldsymbol{e}}_R + E_L \widehat{\boldsymbol{e}}_L, \tag{S1.1}$$

in terms of the right- and left-handed (RH, LH) circular polarization complex unit vectors

$$\widehat{\boldsymbol{e}}_R = \tfrac{1}{\sqrt{2}}(\widehat{\boldsymbol{x}} + i\,\widehat{\boldsymbol{y}}), \qquad \widehat{\boldsymbol{e}}_L = \tfrac{1}{\sqrt{2}}(\widehat{\boldsymbol{x}} - i\,\widehat{\boldsymbol{y}}). \tag{S1.2}$$

Throughout this work, for mathematical convenience, we assume that the field $\boldsymbol{E}$ is normalized, $|E_R|^2 + |E_L|^2 = 1$. Equation (1) of the main text describes $E_R$ and $E_L$ in terms of three angle parameters $\alpha$, $\beta$ and $\gamma$,

$$E_R = \cos\tfrac{\beta}{2}\,e^{i(\gamma-\alpha)/2}, \qquad E_L = \sin\tfrac{\beta}{2}\,e^{i(\gamma+\alpha)/2}.$$

We begin by describing the standard Stokes parameters and Poincaré sphere formalism, and also describe the phase parameter. This leads to a discussion of 3-sphere geometry and the Hopf fibration, which is then related to the parametrisation of equation (S1.1) using the angles $\alpha, \beta, \gamma$.

### 1.1. Optical state parameters

#### 1.1.1. Stokes parameters

The (normalized) *Stokes parameters* $S_1, S_2, S_3$ may be defined from $\boldsymbol{E}$ in terms of Pauli spin matrices in the circular basis,

$$\left.\begin{array}{rcl}
S_1 = \boldsymbol{E}^*\boldsymbol{\sigma}_1\boldsymbol{E} &=& 2\,\mathrm{Re}(E_R^* E_L) = 2\left[(\mathrm{Re}\,E_R)(\mathrm{Re}\,E_L) + (\mathrm{Im}\,E_R)(\mathrm{Im}\,E_L)\right], \\
S_2 = \boldsymbol{E}^*\boldsymbol{\sigma}_2\boldsymbol{E} &=& 2\,\mathrm{Im}(E_R^* E_L) = 2\left[(\mathrm{Re}\,E_R)(\mathrm{Im}\,E_L) - (\mathrm{Im}\,E_R)(\mathrm{Re}\,E_L)\right], \\
S_3 = \boldsymbol{E}^*\boldsymbol{\sigma}_3\boldsymbol{E} &=& |E_R|^2 - |E_L|^2 = (\mathrm{Re}\,E_R)^2 + (\mathrm{Im}\,E_R)^2 - (\mathrm{Re}\,E_L)^2 - (\mathrm{Im}\,E_L)^2,
\end{array}\right\} \tag{S1.3}$$

where $\boldsymbol{\sigma}_1 = \left(\begin{smallmatrix} 0 & 1 \\ 1 & 0 \end{smallmatrix}\right), \boldsymbol{\sigma}_2 = \left(\begin{smallmatrix} 0 & -i \\ i & 0 \end{smallmatrix}\right)$, and $\boldsymbol{\sigma}_3 = \left(\begin{smallmatrix} 1 & 0 \\ 0 & -1 \end{smallmatrix}\right)$. These quadratic combinations of the four fundamental real-valued field quantities, $\mathrm{Re}\,E_R, \mathrm{Im}\,E_R, \mathrm{Re}\,E_L, \mathrm{Im}\,E_L$, only involve information about the polarization state – they are independent of the overall phase, typical of measurable quantum expectation values and classical constants of the motion. With these definitions, it is not difficult to show that $S_1^2 + S_2^2 + S_3^2 = 1$. The resulting normalized 3-vector $(S_1, S_2, S_3)$, the *Stokes vector*, lies on the surface of a unit 2-sphere, the *Poincaré sphere* [1, 2]. Every polarization state (regardless of its phase) corresponds to a point on this sphere: RH circular ($E_L = 0$) to the north pole $(0, 0, +1)$, LH circular ($E_R = 0$) to the south pole $(0, 0, -1)$, linearly polarized states ($S_3 = 0$) to the equator, and so on. More general elliptical polarized states with RH (LH) circulation, with $S_3 > 0$ ($S_3 < 0$) occur in the northern (southern) hemisphere. This parametrisation of elliptic states on the Poincaré sphere is demonstrated in main text Fig. 1.

The three standard Stokes parameters description may be extended by two extra, alternative parameters (mathematically defined in a similar way to $S_1$ and $S_2$) [3, 4] which determine the phase information,

$$\left.\begin{array}{rcl}
S_4 &=& \mathrm{Re}(E_R E_L) = 2\left[(\mathrm{Re}\,E_R)(\mathrm{Re}\,E_L) - (\mathrm{Im}\,E_R)(\mathrm{Im}\,E_L)\right], \\
S_5 &=& \mathrm{Im}(E_R E_L) = 2\left[(\mathrm{Re}\,E_R)(\mathrm{Im}\,E_L) + (\mathrm{Im}\,E_R)(\mathrm{Re}\,E_L)\right].
\end{array}\right\} \tag{S1.4}$$

With $S_3$, these also satisfy the normalization condition $S_3^2 + S_4^2 + S_5^2 = 1$. This means that an alternative sphere similar to the Poincaré sphere can be defined with cartesian coordinates $S_3, S_4, S_5$ instead of $S_1, S_2, S_3$ [3, 4]. We will discuss the significance of this definition below.

### 1.1.2. Polarization ellipse and phase

Apart from the overall normalization, the optical state $\boldsymbol{E}$ is determined by the normalized, complex two-component electric vector $(E_R, E_L)$, or equivalently, the normalized 4-dimensional vector $(\text{Re}\,E_R, \text{Im}\,E_R, \text{Re}\,E_L, \text{Im}\,E_L)$. These parameters, together with the equivalent angles $\alpha, \beta, \gamma$ used in the main text and defined in equation (1), are related to the geometry of the polarization ellipse. The time-dependent, monochromatic electric field $\boldsymbol{E}^{\text{Re}}$ traces out an ellipse in the $xy$-plane in time $t$ with period $2\pi/\omega$:

$$\boldsymbol{E}^{\text{Re}}(\omega t) = \text{Re}\{\boldsymbol{E}\,e^{-i\omega t}\} \tag{S1.5}$$
$$= \tfrac{1}{\sqrt{2}}\left\{[\text{Re}(E_R + E_L)\cos\omega t + \text{Im}(E_R + E_L)\sin\omega t]\,\widehat{\boldsymbol{x}} + [\text{Im}(E_L - E_R)\cos\omega t + \text{Re}(E_R - E_L)\sin\omega t]\,\widehat{\boldsymbol{y}}\right\}.$$

A simple way of visualising this ellipse is in terms of its major axis $\boldsymbol{a}$ and minor axis $\boldsymbol{b}$; the normalization requires $|\boldsymbol{a}|^2 + |\boldsymbol{b}|^2 = 1$ and the axes are perpendicular, $\boldsymbol{a} \cdot \boldsymbol{b} = 0$. The ellipse curve traced out by $\text{Re}\{(\boldsymbol{a} + i\,\boldsymbol{b})\,e^{-i\omega t}\}$ is the same as that traced out by $\boldsymbol{E}^{\text{Re}}(\omega t)$; the only difference is the starting point on the ellipse, which is related to the phase parameter $\gamma$.

The phase of $\boldsymbol{E}$ is related to the overall phase factor at $t = 0$ multiplying the complex combination of major and minor axis vectors at time $t = 0$, i.e. $\boldsymbol{E} = e^{i\gamma/2}(\boldsymbol{a} + i\,\boldsymbol{b})$. From the discussion above, this implies that $\boldsymbol{E}^{\text{Re}}(\omega t = \gamma/2) = \boldsymbol{a}$. Thus the effective phase evolves with time. It is necessary to be careful with this definition, as described in the following.

The major and minor axes correspond, with respect to $t$, to maxima and minima of

$$|\boldsymbol{E}^{\text{Re}}(\omega t)|^2 = \tfrac{1}{2}[1 + S_4\cos(2\omega t) + S_5\sin(2\omega t)], \tag{S1.6}$$

which has been simplified using the alternative Stokes parameters (S1.4).

This is a maximum at $\omega t_{\max} = \gamma_+$ and a minimum at $\omega t_{\min} = \gamma_-$, where

$$\gamma_+ \equiv \tfrac{1}{2}\arctan(S_4, S_5) = \tfrac{1}{2}\arg(E_R E_L), \qquad \gamma_- \equiv \tfrac{1}{2}\arctan(-S_4, -S_5). \tag{S1.7}$$

Since arctan is a periodic function with period $\pi$, there is another maximum/minimum at $\gamma_\pm + \pi$. We therefore have

$$\boldsymbol{a} = \boldsymbol{E}^{\text{Re}}(\gamma_+), \qquad \boldsymbol{b} = \boldsymbol{E}^{\text{Re}}(\gamma_-). \tag{S1.8}$$

In fact, these give a simple expression for the original complex vector $\boldsymbol{E}$,

$$\boldsymbol{E} = (\boldsymbol{a} + i\,\boldsymbol{b})\,e^{i\gamma_+}, \tag{S1.9}$$

where $\gamma_+$ determines the phase factor multiplying the complex vector of major and minor axes, with a $\pi$ ambiguity. The definition of the phase parameter $\gamma$ correctly over the $4\pi$ range, i.e. as $2\gamma_+$ or $2\gamma_+ \pm 2\pi$, requires some care, as described below. With this in mind, we discuss each of the angles $\alpha, \beta, \gamma$ used in the main text and defined in main text equation (1).

### 1.1.3. Polarization azimuth $\alpha$ and ellipticity $\beta$

The angle $\alpha$ is defined as the azimuth angle on the Poincaré sphere,

$$\alpha = \arctan(S_1, S_2) = \arg(E_R^* E_L). \tag{S1.10}$$

As shown in Supplementary Fig. 1, $\alpha/2$ is the angle between the polarization ellipse major axis and the $x$-direction, i.e. $\boldsymbol{a} = |\boldsymbol{a}|\left(\cos(\tfrac{1}{2}\alpha)\widehat{\boldsymbol{x}} + \sin(\tfrac{1}{2}\alpha)\widehat{\boldsymbol{y}}\right)$. This follows from equation (S1.8), after simplification and substituting the expression for the maximum $\gamma_+$ from equation (S1.7) into (S1.5).

This half angle cannot distinguish between $\pm\boldsymbol{a}$: the orientation angle $\alpha/2$ of the major axis of the ellipse is in fact a *director*, only defined over a $\pi$ range, say $-\pi/2 < \alpha/2 \leq \pi/2$, consistent with our definition of the range $-\pi < \alpha \leq \pi$. This half-angle nature of $\alpha$ implies that the azimuth angle on the Poincaré sphere changes twice as quickly as the ellipse orientation angle in real space, and has been subject to much comment in the literature. The factor of $1/2$ on $\alpha$ in equation (1) is naturally associated with spinorial parametrisation. Its range from $-\pi$ to $\pi$ is consistent with its definition as the argument of the product $E_R^* E_L$.

The angle $\beta$ is the polar angle on the Poincaré sphere,

$$\beta = \arccos S_3, \qquad \sin\beta = \sqrt{S_1^2 + S_2^2} = \sqrt{S_4^2 + S_5^2}. \tag{S1.11}$$

It determines the shape and the handedness of the ellipse, and it can also be shown that

$$\cos\beta = 2\hat{z}\cdot(\boldsymbol{a}\times\boldsymbol{b}) = 2|\boldsymbol{a}||\boldsymbol{b}|\operatorname{sign} S_3. \tag{S1.12}$$

Therefore, $\cos\beta$ is proportional to the area of the bounding rectangle of the ellipse, with the normalization condition $|\boldsymbol{a}|^2 + |\boldsymbol{b}|^2 = 1$ corresponding to fixing the main diagonal of the bounding rectangle, as shown in Supplementary Fig. 1. Since $-1 \le S_3 \le 1$, the range of $\beta$ is from 0 to $\pi$.

### 1.1.4. Optical phase and $\gamma$

The definition above of the phases $\gamma_\pm$ at the maxima and minima ignored the subtlety that we did not distinguish between $+\boldsymbol{a}$ and $-\boldsymbol{a}$. In our previous discussion, we saw that this only determines the ellipse major axis as a director $\pm\boldsymbol{a}$, and the azimuth parameter $\alpha$ (with $-\pi < \alpha \le \pi$) is twice the angle between the major axis and $\hat{\boldsymbol{x}}$.

In the definition of $\gamma_+$ in equation (S1.7), the choice of $\gamma_+ = \frac{1}{2}\arg(E_R E_L)$ rather than $\frac{1}{2}\arg(E_R E_L) + \pi$ was arbitrary, and is defined $-\pi/2 < \gamma_+ \le \pi/2$. This choice represents the phase difference between the point on $\boldsymbol{E}^{\mathrm{Re}}(0)$ on the ellipse and the closer of the two major axis vectors $\pm\boldsymbol{a}$. As the phase cycles around the whole ellipse, which one of $\pm\boldsymbol{a}$ is closer to $\boldsymbol{E}^{\mathrm{Re}}(\gamma/2)$ jumps, as $\boldsymbol{E}^{\mathrm{Re}}$ crosses the minor axis $\pm\boldsymbol{b}$ when $\gamma_+ = \pm\pi/2$.

To determine the phase value unambiguously, it is necessary to fix a choice of major axis vector $\pm\boldsymbol{a}$ for each polarization state, say $+\boldsymbol{a}$. This allows the phase position of $\boldsymbol{E}^{\mathrm{Re}}(0)$ to be determined with respect to $\boldsymbol{a}$, which extends the range of $\gamma_+$ to a full cycle, $-\pi < \gamma_+ \le \pi$.

Just as the angles $\alpha$ and $\beta$ are defined as the half-angles of the natural physical angle, so is the phase parameter $\gamma$ twice the value of the "physical" phase angle, $\gamma = 2\gamma_+$. This implies that $\gamma$ varies over a $4\pi$ range, $-2\pi < \gamma \le 2\pi$, with the general definition

$$\gamma = \arg E_R + \arg E_L. \tag{S1.13}$$

In practice significant care is needed numerically to determine the correct branch cut of $\gamma$ when $\boldsymbol{E}^{\mathrm{Re}}$ crosses $-\boldsymbol{a}$. A numerical value for $\gamma$ can be determined nonlocally for instance by tracing the evolution of $\boldsymbol{E}^{\mathrm{Re}}$ with respect to the chosen sign of the major axis vector.

The normalized polarization state $(E_R, E_L)$ resembles a 2-component spinor with phase. These are often used to represent spatial orientations and rotations (e.g. using the Cayley-Klein parametrisation of Euler angles [5]), parametrising the special orthogonal group SO(3). On the other hand, the full state of light, being parametrised by the 3-sphere, corresponds to a point in the special unitary group SU(2), which is the double cover of SO(3) (i.e. for every point in SO(3) there are two in SU(2)). This double covering, in the $\alpha, \beta, \gamma$ parametrisation, corresponds to the $2\pi$ versus $4\pi$ range of $\gamma$ [6].

### 1.1.5. The Poincaré sphere and its projections

The approach to polarization and the optical state $(E_R, E_L)$ in the circular bases is very reminiscent of the mathematical spinor formalism. This is most familiar in the representation of the state of a quantum spin 1/2 particle, where the spin direction in 3D real space is represented by a point on the Bloch sphere.

Physically, a monochromatic electric field has spin 1. Since we only consider a 2-component transverse field, the complex field $(E_R, E_L)$ mathematically represents a normalized complex spinor. The optical state is not truly a spin 1/2 system, therefore, and so $(E_R, E_L)$ is properly a *pseudospinor*. The corresponding Poincaré sphere is defined in an abstract 3D space, unlike the Bloch sphere, whose directions correspond to those in 3D real space.

An alternative representation of polarization states is via the complex ratio

$$\Psi = \frac{E_L}{E_R} = \tan(\beta/2)\,e^{i\alpha} = \rho(\beta)\,e^{i\alpha}, \tag{S1.14}$$

which defines a point in the complex plane with modulus (radius) $\tan(\beta/2)$ and azimuth $\alpha$. This point corresponds to the *stereographic projection* of the Stokes vector, centred at the north pole, and including the point in the plane "at infinity" ($\beta = \pi$, i.e. the south pole), the plane is topologically equivalent to the sphere. This is a direct illustration of *one-point compactification*: a euclidean plane, with a single point $\infty$ adjoined, becomes, topologically, identical to a sphere. This procedure can be generalised directly to euclidean $n$-space and the $n$-sphere.

In principle, any positive monotonic function $\rho(\beta)$ can be used as a radial coordinate to map the 2-sphere to the plane; when $\rho(\beta = \pi)$ is finite, the map is in a disk of finite radius, and all points on the circumference $\rho(\pi)$ correspond to the south pole. In the example of stereographic coordinates above, $\rho = \tan(\beta/2)$; this projection from the sphere to the plane is conformal, preserving angles. Another important example is $\rho(\beta) = 2\cos(\beta)$, *Lambert equal area projection*, where areas on the plane correspond to areas on the sphere. Areas are preserved here even though the south pole $\beta = \pi$ projects to a circle (radius 2).

The stereographic representation (S1.14) is based on a *north pole projection*, which associates the north pole $\beta = 0$ with the origin of the plane, and the south pole $\beta = \pi$ with $\infty$. With this choice, positive orientation (the local right-handed sense on the sphere) is consistent with the orientation of the increase of the sphere azimuth coordinate $\alpha$. We could instead have chosen *south pole projection* where $\rho = \cot(\beta/2)$, which associates the south pole with the origin and the north pole with $\infty$. For this projection to preserve orientation, positive orientation is associated with a *decrease* in sphere azimuth angle $\alpha$, so the complex coordinate associated with orientation-preserving south pole projection is $\cot(\beta/2)\,e^{-i\alpha}$. This is important for the analogous argument for the 3-sphere.

## 1.2. The hypersphere and the Hopf fibration

### 1.2.1. Parametrising the hypersphere

Many of the details discussed above about 2-sphere geometry, topology and parametrisations generalise directly to the 3-sphere (hypersphere) [7, 8]. For instance, we can consider 4-dimensional (4D) space with cartesian coordinates $W, Z, X, Y$. The reason for the non-standard ordering is as follows.

The 3-sphere is defined by all possible 4D unit vectors, represented using overarrow $\vec{n} = (n_W, n_Z, n_X, n_Y)$. Points on the 3-sphere can be parametrised by hyperspherical polar angles $\omega, \theta, \varphi$ with $0 \leq \omega, \theta \leq \pi$, $-\pi < \varphi \leq \pi$ in the following way,

$$\left.\begin{aligned}
n_W &= \cos\omega, \\
n_Z &= \sin\omega\cos\theta, \\
n_X &= \sin\omega\sin\theta\cos\varphi, \\
n_Y &= \sin\omega\sin\theta\sin\varphi.
\end{aligned}\right\} \tag{S1.15}$$

Clearly these satisfy $n_W^2 + n_Z^2 + n_X^2 + n_Y^2 = 1$. (Note that the 4-space azimuthal angle $\varphi$ is different from the 3D real space azimuthal angle $\phi$ used later.)

Features of the topology of the 3-sphere can also be discerned from this parametrisation, by relating the $XYZ$ coordinates to the 3D cartesian coordinates. $n_X$, $n_Y$, $n_Z$ resemble the coordinates of a 2-sphere, parametrised by spherical polar angles $\theta$ and $\varphi$ as normal, and with radial coordinate $\sin\omega$: all these spheres enclose a central point (at which $\omega = 0$), and extend out to a sphere of unit radius at $\omega = \pi$. In this embedding, the boundary sphere corresponds to the single point $\omega = \pi$ in the 3-sphere (similar to the embedding of the 2-sphere in the 2-plane as discussed above). This solid ball in 3D is therefore a projection of the 3-sphere into euclidean 3D space, for which $\sin\omega = \rho(\omega)$ as defined in Sec. 1.1, related to the quaternion representation of rotations (where the angle of rotation is $2\omega$) [9, 10].

Just as for the 2-sphere, we can consider different "ball" projections of the 3-sphere into 3D euclidean space by different choices of $\rho(\omega)$. For instance, stereographic projection to $X, Y, Z$ coordinates is achieved via $\rho(\omega) = \tan\frac{1}{2}\omega$. Explicitly, this gives

$$n_X = \frac{2X}{1 + X^2 + Y^2 + Z^2}, \quad n_Y = \frac{2Y}{1 + X^2 + Y^2 + Z^2}, \quad n_Z = \frac{2Z}{1 + X^2 + Y^2 + Z^2}, \quad n_W = \frac{1 - X^2 - Y^2 - Z^2}{1 + X^2 + Y^2 + Z^2}. \tag{S1.16}$$

As in 2D stereographic projection (S1.14), the 3-sphere point $\omega = \pi$ corresponds in euclidean 3D space to the point "$\infty$". We can also write complex coordinates $U, V$ for the 3-sphere,

$$\left.\begin{aligned}
U(X,Y,Z) &= n_W + i\,n_Z &= (1 - X^2 - Y^2 - Z^2 + 2i\,Z)/(1 + X^2 + Y^2 + Z^2), \\
V(X,Y,Z) &= n_X + i\,n_Y &= 2(X + iY)/(1 + X^2 + Y^2 + Z^2),
\end{aligned}\right\} \tag{S1.17}$$

where in each case the second equality corresponds to stereographic projection to euclidean space (with the point with $\omega = \pi$, $n_W = -1$ at infinity), and $|U|^2 + |V|^2 = 1$. The modulus and phase of $U$ are shown in Supplementary Fig. 2a; as described below, with $\varphi$ these correspond to toroidal coordinates in euclidean 3-space.

Stereographic projection provides a very natural visualisation of the hypersphere in terms of familiar, infinite 3D euclidean space. Yet other projections of the 3-sphere into euclidean 3-space are possible [9, 10]. The representations of the optical hypersphere in main text Fig. 4, are chosen to preserve 3D volume, analogous to Lambert equal area projection, for which the radius function is $\rho(\omega) = [\frac{3}{4}(2\omega - \sin 2\omega)]^{1/3}$ [9].

### 1.2.2. Hopf fibration

We have considered the polar angle parametrisation of the hypersphere using $\omega, \theta, \varphi$ by analogy with the polar angles of the 2-sphere. They are not the only natural parameters of the 3-sphere, and in fact are not the natural physical parametrisation of the optical hypersphere.

We identify the complex scalar $U$ with the RH component of the optical field, i.e. $U = E_R$, and $V$ with the LH component, $V = E_L$; this is possible since both pairs of complex scalars are normalized, $1 = |U|^2 + |V|^2 = |E_R|^2 + |E_L|^2$. This identification gives the optical state parameters $\alpha, \beta, \gamma$ as the parameters of the Hopf fibration in the optical hypersphere, described in the following.

Firstly, the identification of $U$ and $V$ with $E_R$ and $E_L$ allows us to find the relationship between $\alpha, \beta, \gamma$ and $\theta, \varphi, \omega$,

$$\left.\begin{aligned} E_R &= \cos(\beta/2)\, e^{i(\gamma-\alpha)/2} &= \cos\omega + i\sin\omega\cos\theta, \\ E_L &= \sin(\beta/2)\, e^{i(\gamma+\alpha)/2} &= \sin\omega\sin\theta\, e^{i\varphi}. \end{aligned}\right\} \tag{S1.18}$$

The unit 4-vector $\vec{n}$ is identified with the 4-vector of real and imaginary parts of $E_R$ and $E_L$ considered earlier,

$$\begin{aligned} \vec{n} &= (\operatorname{Re} E_R, \operatorname{Im} E_R, \operatorname{Re} E_L, \operatorname{Im} E_L) \\ &= \left(\cos\tfrac{1}{2}\beta\cos\tfrac{1}{2}[\gamma-\alpha], \cos\tfrac{1}{2}\beta\sin\tfrac{1}{2}[\gamma-\alpha], \sin\tfrac{1}{2}\beta\cos\tfrac{1}{2}[\gamma+\alpha], \sin\tfrac{1}{2}\beta\sin\tfrac{1}{2}[\gamma+\alpha]\right) \\ &= (\cos\omega, \sin\omega\cos\theta, \sin\omega\sin\theta\cos\varphi, \sin\omega\sin\theta\sin\varphi)\,. \end{aligned} \tag{S1.19}$$

We now interpret $\alpha, \beta$ and $\gamma$ in the 3-sphere, represented by euclidean 3D space under stereographic projection. This is possible since the Hopf fibration of the 3-sphere can in fact be constructed directly from *toroidal coordinates* [11–13] in euclidean 3-space, considered as the stereographic projection of the 3-sphere.

Consider the complex coordinates $U, V$ as complex scalar functions of 3D euclidean space. $U$ has a circular nodal line (phase singularity) of unit radius in the $XY$-plane centred on the origin, which we call the *reference circle* (anticipating the use of toroidal coordinates below). This is threaded by a nodal line in $V$ along the $Z$-axis (which corresponds to a circle in the 3-sphere, with one point mapping to $\infty$ in euclidean 3-space), which we call the *axial line*. As a stereographic representation of the optical hypersphere, the states of LH polarization ($\beta = 0$) occur on the reference circle, and the states of RH polarization ($\beta = \pi$) on the axial line (this includes the point at $\infty$, closing topologically into a loop). The phase varies smoothly along these lines. Representing the RH and LH states of polarization, the axial line and reference circle are special, as curves in the 3D optical hypersphere, corresponding to the points of circular polarization at the poles of the 2D Poincaré sphere.

We now construct the Hopf structure using the fact that $U$ and $V$ are naturally related to toroidal coordinates $\varphi, \sigma, \tau$ in $X, Y, Z$ space [11–13]. The toroidal coordinate $\varphi$ is the familiar azimuth angle of spherical and cylindrical polar coordinates,

$$\varphi = \arctan(X, Y) = \arg V = \tfrac{1}{2}(\gamma + \alpha), \tag{S1.20}$$

from equations (S1.17) and (S1.18). We use the standard convention of increase of $\varphi$, but note it needs to be reversed when considering 3D orientation. The coordinate $\sigma$ is the poloidal angle about the reference circle $U = 0$ (also see Supplementary Fig. 2a), and in fact, also from equations (S1.18) and (S1.17),

$$\sigma = \arctan(X^2 + Y^2 + Z^2 - 1, 2\,i\,Z) = \arg U = \tfrac{1}{2}(\gamma - \alpha). \tag{S1.21}$$

On the punctured plane $Z = 0, R > 0$, $\sigma = 0$, and $\sigma$ increases around the reference circle in a RH sense in a plane of fixed $\varphi$, as in Supplementary Fig. 2a. The final toroidal coordinate, $\tau$, is defined over the range $0 \leq \tau < \infty$, by

$$\cosh\tau = \frac{X^2 + Y^2 + Z^2 + 1}{\sqrt{(X^2 + Y^2 + Z^2)^2 + 1 - 2(X^2 + Y^2 - Z^2)}} = |U|^{-1} = \sec\tfrac{1}{2}\beta = \sqrt{\frac{2}{1 + \cos\beta}}, \tag{S1.22}$$

which is simply a monotonic function of $\beta$. Surfaces of constant $\tau$ are tori centred on the axial line which enclose the reference circle, with major radius $\coth\tau$ and minor radius $\operatorname{cosech}\tau$. $\tau = 0$ on the axial line, with $\tau \to \infty$ on the reference circle.

Now, the azimuthal and poloidal angles $\varphi$ and $\sigma$ are angles winding around the tori of constant $\tau$ in flat planes; their sum and difference, $\alpha = \varphi - \sigma$ and $\gamma = \varphi + \sigma$, both wind around the tori as $(1, 1)$ torus curves. Loci of constant $\beta$ and $\alpha$, therefore give curves—in fact Villarceau circles—which wrap around both nodal lines in $U, V$ as $\gamma$. The resulting structure of lines on which $\alpha, \beta$ are constant define the *Hopf fibration*, as in Supplementary Fig. 3. Position on these Hopf circles is parametrised by $\gamma$ on its $4\pi$ cycle.

The Hopf fibration represents the 3-sphere as a fiber bundle, whose fibers are the Villarceau circles for fixed $\alpha, \beta$ along which $\gamma$ varies. The parameters $\beta, \alpha$ themselves are determined by points on the Poincaré 2-sphere (the *base space* of the fibration). The loci of constant $\beta$ (i.e. constant $S_3$) correspond to constant $\tau$, i.e. tori, around which are the nodal line loops corresponding to $\beta = 0, \pi$.

The choice of $\alpha, \beta$ parametrising the base space and $\gamma$ parametrising the fibers is only one possibility; it is also possible to parametrise the fibers with $\alpha$, in which case the base space is the sphere parametrised by $\alpha, \gamma$ (with appropriate modification to the ranges). The sphere of this fibration is via the Hopf map of the alternative Stokes parameters $S_3, S_4, S_5$ introduced in equation (S1.4).

### 1.2.3. 3-sphere volume

The unit 3-sphere has total volume (hypersurface area) $2\pi^2$, as we now discuss. The normalized volume element $\mathrm{d}^3\Omega$ of the hypersphere can be derived from the hypersurface area of the 3-sphere in 4D space, in terms of either set of angle coordinates $\theta, \varphi, \omega$, or $\alpha, \beta, \gamma$. Some care is needed to ensure the proper sense of orientation on this 3-manifold; we define this in terms of the usual orientation of euclidean 3D space for stereographic coordinates, which corresponds to ordering the angles $\omega, \theta, \varphi$.

In terms of $\omega, \theta, \varphi$, the normalized volume element (dividing by the hypersphere volume $2\pi^2$) is

$$\mathrm{d}^3\Omega = \frac{1}{2\pi^2} \det[\,\vec{n}, \partial_\omega \vec{n}, \partial_\theta \vec{n}, \partial_\varphi \vec{n}\,]\,\mathrm{d}\omega\,\mathrm{d}\theta\,\mathrm{d}\varphi = \frac{1}{2\pi^2} \sin^2 \omega \sin \theta\,\mathrm{d}\omega\,\mathrm{d}\theta\,\mathrm{d}\varphi, \tag{S1.23}$$

where $\det[\,\vec{n}, \partial_\omega \vec{n}, \partial_\theta \vec{n}, \partial_\varphi \vec{n}\,]$ denotes the matrix of column vector partial derivatives of $\vec{n}$, calculated using equation (S1.19). This determinant plays the role, in 4D space, of the scalar triple product in 3D. With this normalization, the integral of these angles over the 3-sphere with ranges $0 \leq \omega, \theta \leq \pi$, $-\pi < \varphi \leq \pi$ indeed gives unity.

We can apply this procedure to the optical state parameters $\alpha, \beta, \gamma$. In order to preserve the sign of the determinant (i.e. the 3-sphere orientation), it is necessary to use the ordering $\gamma, \beta, \alpha$ (or a cyclic permutation of them), i.e.

$$\mathrm{d}^3\Omega = \frac{1}{2\pi^2} \det[\,\vec{n}, \partial_\gamma \vec{n}, \partial_\beta \vec{n}, \partial_\alpha \vec{n}\,]\,\mathrm{d}\gamma\,\mathrm{d}\beta\,\mathrm{d}\alpha = \frac{1}{16\pi^2} \sin \beta\,\mathrm{d}\gamma\,\mathrm{d}\beta\,\mathrm{d}\alpha = \frac{1}{16\pi^2}\,\mathrm{d}\alpha\,\mathrm{d}(\cos\beta)\,\mathrm{d}\gamma. \tag{S1.24}$$

In the last equality, $\beta$ was replaced by $\cos\beta$; the volume element in terms of $\alpha, \cos\beta, \gamma$ has no prefactor, and is positive in lexicographical ordering (since $\cos\beta$ increases as $\beta$ decreases).

### Supplementary Note 2: Designing the optical skyrmionic hopfion

In this note we summarise the theoretical model of the skyrmionic hopfion field. This field configuration captures the Hopf fibration structure in real space, realising all phases and polarizations within the propagation volume. We construct solutions to the (paraxial) optical wave equation that show the desired topology, then we describe how we select the optimal superposition of Laguerre-Gauss modes representing a physically realisable optical skyrmionic hopfion.

Such a field is designed by creating RH and LH fields, as functions of real 3D space using cartesian coordinates $x, y, z$ or cylindrical coordinates $R, \phi, z$, i.e. $E_\mathrm{R}(x, y, z)$ and $E_\mathrm{L}(x, y, z)$. We introduce unnormalized forms of the field components $\psi_\mathrm{R}(x, y, z), \psi_\mathrm{L}(x, y, z)$, such that

$$E_\mathrm{R} = \frac{\psi_\mathrm{R}}{\sqrt{|\psi_\mathrm{R}|^2 + |\psi_\mathrm{L}|^2}}, \qquad E_\mathrm{L} = \frac{\psi_\mathrm{L}}{\sqrt{|\psi_\mathrm{R}|^2 + |\psi_\mathrm{L}|^2}}. \tag{S2.1}$$

The scalar fields $\psi = \psi_\mathrm{R}, \psi_\mathrm{L}$ are solutions of the paraxial wave equation $\nabla_\perp^2 \psi + 2\,i\,k\partial_z \psi = 0$, where $\nabla_\perp^2 = \partial_x^2 + \partial_y^2$ is the 2D transverse laplacian, $k = 2\pi/\lambda$ is the wavenumber; these are good representations of scalar components of collimated laser beams [14]. For future consideration, we define the phases of the scalar components, $\chi_\mathrm{R} \equiv \arg \psi_\mathrm{R}$, and $\chi_\mathrm{L} \equiv \arg \psi_\mathrm{L}$.

As described in the main text and methods, the skyrmionic hopfion field is designed by a superposition of Laguerre-Gaussian (LG) beams, with the RH component $\psi_\mathrm{R}$ being zero on a circle in the focal plane, and $\psi_\mathrm{R}$ zero on the beam axis. The zeros of the circular components correspond to C lines, along which the state of polarization is circular [15]. In the full polarization field, these correspond to a RH C line along the axis, and a LH C line circle. The design of the optical skyrmionic hopfion field is thus based on the principle that C lines are the skeleton of the complex optical polarization field, although careful tuning of the superposition is necessary to guarantee the covering of the optical hypersphere within the focal volume.

## 2.1. Laguerre-Gauss modes

Laguerre-Gauss (LG) modes are a convenient set of basis functions for solutions of the paraxial wave equation. The LG modes are naturally expressed in cylindrical coordinates $R, \phi, z$ by [14, 16]

$$\mathrm{LG}^{3D}_{\ell,p}(R, \phi, z; w) = \sqrt{\frac{p!}{\pi(|\ell| + p)!}} \frac{(R/w)^{|\ell|} e^{i\ell\phi}}{w} \frac{(1 - iz/z_R)^p}{(1 + iz/z_R)^{p+|\ell|+1}} \mathrm{L}_p^{|\ell|}\left(\frac{(R/w)^2}{1 + (z/z_R)^2}\right) \exp\left(-\frac{(R/w)^2}{2(1 + iz/z_R)}\right) \quad \text{(S2.2)}$$

$$= \sqrt{\frac{p!}{\pi(|\ell| + p)!}} \frac{R^{|\ell|} e^{i\ell\phi}}{w^{|\ell|+1}} \frac{e^{-R^2/w^2(1 + z^2/z_R^2)}}{(1 + (\frac{z}{z_R})^2)^{|\ell|+1}} \mathrm{L}_p^{|\ell|}\left(\frac{(R/w)^2}{1 + (\frac{z}{z_R})^2}\right) e^{i(|\ell|+2p+1)\arctan(z/z_R)} e^{i \cdot R^2 z z_R/w^2(z_R^2 + z^2)}. \quad \text{(S2.3)}$$

Here, $w$ is the width of the gaussian envelope setting the transverse size of the beam, and $z_R = kw^2$, the Rayleigh range of the gaussian beam, determines the longitudinal lengthscale. $\mathrm{L}_p^{|\ell|}$ denotes an associated Laguerre polynomial [12, 14]. The positive or negative (or zero) integer $\ell$ is the azimuthal quantum number; when $\ell \neq 0$, there is an optical vortex along the axis of topological strength $\ell$. The nonnegative integer $p$ is the radial order, corresponding to the order of the Laguerre polynomial, giving the number of nodal rings concentric to the axis in a typical transverse plane. The term $\mathrm{LG}_{\ell,p}$ in the main text corresponds to $\mathrm{LG}^{3D}_{\ell,p}(R, \phi, z; w)$ for some fixed value of $w$ (and of course $k$).

Although the equation (S2.2) is more useful for calculations, the form (S2.3), in which complex factors have had their modulus and phase separated, is easier to interpret physically. In particular, the LG beam has two important $z$-dependent phase factors: the *Gouy phase* $e^{i(|\ell|+2p+1)\arctan(z/z_R)}$, and the wavefront curvature $e^{i R^2 z z_R/w^2(z_R^2+z^2)}$. Neither of these contribute in the waist plane $z = 0$.

For each value of $z$ (i.e. transverse plane), the LG modes are an orthonormal basis for 2D complex amplitude distributions. The simplest such plane is the focal plane $z = 0$, for which

$$\mathrm{LG}^{2D}_{\ell,p}(R, \phi; w) \equiv \mathrm{LG}^{3D}_{\ell,p}(R, \phi, z = 0; w) = \sqrt{\frac{p!}{\pi(|\ell| + p)!}} \frac{(R/w)^{|\ell|} e^{i\ell\phi}}{w} \mathrm{L}_p^{|\ell|}\left(\frac{R^2}{w^2}\right) \exp\left(-\frac{R^2}{2w^2}\right). \quad \text{(S2.4)}$$

As the fields are synthesised by 2D holograms in a beam's transverse plane, we approach the 3D field design primarily in terms of the field distributions in the 2D field [17]. Following standard linear space theory, any scalar solution $\psi(R, \phi, z)$ of the paraxial wave equation can be written as a superposition

$$\psi(R, \phi, z) = \sum_{\ell,p} c_{\ell,p} \mathrm{LG}^{3D}_{\ell,p}(R, \phi, z; w), \quad \text{(S2.5)}$$

where the generally complex coefficients are found from the 2D inner product integral

$$c_{\ell,p} = \int_0^\infty \int_0^{2\pi} \psi(R, \phi, z = 0) \mathrm{LG}^{2D*}_{\ell,p}(R, \phi; w) R \, d\phi \, dR. \quad \text{(S2.6)}$$

## 2.2. Mapping from 3D real space to the optical hypersphere

Any transverse optical field defines a mapping from 3D real space, parametrised by cartesian $x, y, z$ or cylindrical polar $R, \phi, z$, to the optical hypersphere, which is parametrised in terms of $X, Y, Z, \omega, \theta, \varphi$, or $\alpha, \beta, \gamma$ as discussed in Sec. 1.2, each of which are different ways of expressing the complex components $E_R = U, E_L = V$. In this section we will be careful distinguish the coordinates in the target space (i.e. the optical hypersphere), and the topological mapping to them from real space, as the notation of the complex fields $E_R(x, y, z)$ and $E_L(x, y, z)$ suggests both.

The most direct such map would be an optical field which maps each point of real space to the corresponding point in the stereographic projection of the optical hypersphere, i.e. $(x, y, z) \longrightarrow (X, Y, Z)$. Under such a map, each point of real space would correspond to a unique optical state (including the limiting point "at infinity"), wrapping around the optical hypersphere exactly once. However, such a mapping is not compatible with a physically realisable optical field, satisfying the paraxial equation.

As a step towards this, we define the following real-space counterparts to $U$ and $V$ of equation (S1.17)

$$u(x, y, z) = \frac{x^2 + y^2 + z^2 - 1 + 2iz}{x^2 + y^2 + z^2 + 1} = \frac{R^2 + z^2 - 1 + 2iz}{R^2 + z^2 + 1}, \qquad v(x, y, z) = \frac{2(x + iy)}{x^2 + y^2 + z^2 + 1} = \frac{2R e^{i\phi}}{R^2 + z^2 + 1}. \quad \text{(S2.7)}$$

In the correspondence between $x, y, z$ and $X, Y, Z$, $\mathrm{Re}\, u$ corresponds to $-\mathrm{Im}\, U$, as we explain below.

If we could realise a field with $E_\mathrm{R} = u(x, y, z)$ and $E_\mathrm{L} = v(x, y, z)$, we would realise a reflection of the skyrmionic hopfion (reflected in the $n_W = 0$ plane), but as stated above, this is not possible physically with solutions of the paraxial wave equation. The vortex function $\psi_1(x, y, z) = 2(x \pm \mathrm{i}\, y)$, formally satisfies the paraxial equation, and can be taken as a local expansion of a physically realisable $\psi_\mathrm{L}$ giving the right form agreeing with the numerator of $v$. Furthermore, the function $\psi_2(x, y, z) = x^2 + y^2 - 1 + 2\mathrm{i}\, z/k$ also satisfies the paraxial wave equation, so can represent a quadratic expansion of a physical field [18]. It has a circular vortex line around the axis in the $z = 0$ plane, and in fact agrees with the numerator of $u$ when $z$ is small.

The field $\psi_1$ satisfies the paraxial wave equation, although its conjugate $\psi^*$ does not; $u$ was defined in equation (S2.7) to approximately agree with $\psi_1$, in spite of it mapping to a reflection of the hypersphere. Positive orientation can be restored to our mapping by choosing to generate $\psi_2^*$ in the other field, rather than $\psi_2$, i.e. a beam with a negative, rather than positive, axial vortex. This guarantees that the covering of the hypersphere by the real space beam has a positive sense, and the net mapping degree—the essential Skyrme number—should be close to $+1$.

Therefore beams with $\psi_\mathrm{L} = \psi_1^*$ and $\psi_\mathrm{R} = \psi_2$, realise the correct topological mapping from real space to the hypersphere. Although not perfect realisations of the $1 - 1$ mapping between the two 3D spaces, the physically realisable optical field we will describe is a smooth deformation of these, homotopically equivalent to the $\Pi_3$ mapping around the hypersphere, except a small neighbourhood near $\infty$ as we truncate within a finite experimental volume.

## 2.3. Topological design via C line skeleton

We thus design the structured fields $E_\mathrm{R}(x, y, z)$ and $E_\mathrm{L}(x, y, z)$ by matching the topology of $\psi_\mathrm{R}(x, y, z)$ and $\psi_\mathrm{L}(x, y, z)$ to the complex fields $\psi_2(x, y, z)$ and $\psi_1^*(x, y, z)$, from Sec. 2.2 as closely as possible in the focal volume. The main design principle for the topology of the skyrmionic hopfion is therefore a vortex loop around the beam axis in the RH component (which corresponds to a LH C line loop in the full field), threaded by a straight vortex line along the $z$-axis in the LH component (corresponding to a RH C line).

As with other topological singular configurations designed in 3D [17, 19, 20], each of the RH and LH components is considered as a superposition of LG modes. The simplest way to generate a vortex ring in $\psi_\mathrm{R}$ is in the focal plane, by a superposition of $\mathrm{LG}_{0,0}$ and $\mathrm{LG}_{0,1}$. Thus, for this component all coefficients (S2.6) are zero except $c_{0,0}$ and $c_{0,1}$, which we redefine them in terms of new parameters $a, b$ as:

$$c_{0,0} = -a + b, \qquad c_{01} = -b.$$ (S2.8)

Therefore, the RH field is

$$\psi_\mathrm{R} = \frac{\mathrm{e}^{-R^2/2w^2(1+\mathrm{i}\, z/z_\mathrm{R})}}{\sqrt{\pi} w (1 + \mathrm{i}\, z/z_\mathrm{R})^3} \left[ b(R^2/w^2 + 2\mathrm{i}\, z/z_\mathrm{R}(1 + \mathrm{i}\, z/z_\mathrm{R}) - a(1 + \mathrm{i}\, z/z_\mathrm{R})^2 \right].$$ (S2.9)

Here, the coefficients $a$ and $b$ tailor the radius of the vortex ring, the shape of the vortex core, and the related intensity. The superposition is chosen so the optical vortex ring has radius $R = R_0 = \sqrt{a/b} w$ $(a, b > 0)$ in the focal plane $z = 0$. In order to get a positive Skyrme number, it is necessary to use a negatively signed axial vortex in $\psi_\mathrm{L}$. This is equivalent to matching the topology of $v^*(x, y, z)$, necessary to guarantee the mapping preserves the sign of 3D volume is preserved; more details are in Supplementary Note 4. For $\psi_\mathrm{L}$, therefore, all $c_{\ell,p}$ are zero except for $c_{-1,0}$, which we choose to be

$$c_{-1,0} = 2c,$$ (S2.10)

involving the new constant $c > 0$. Thus, the LH component is given by

$$\psi_\mathrm{L} = \frac{\mathrm{e}^{-R^2/2w^2(1+\mathrm{i}\, z/z_\mathrm{R})}}{\sqrt{\pi} w (1 + \mathrm{i}\, z/z_\mathrm{R})^3} \left[ 2cR/w\, \mathrm{e}^{-\mathrm{i}\phi}(1 + \mathrm{i}\, z/z_\mathrm{R}) \right].$$ (S2.11)

This expression outside the square brackets has the same nonzero prefactor as the expression for $\psi_\mathrm{R}$. From the perspective of the polarization, all that matters are the parts in the square brackets of equations (S2.9) and (S2.11).

Given fixed values of $k$ and $w$, the optical skyrmionic hopfion forms in a parameter range of $a, b, c$. The criteria by which the parameters are optimised for experiments are reported in the Methods. When $a = b = c = 1$, the terms in the bracket become $R^2/w^2 - 1$ and $2R/w\, \mathrm{e}^{-\mathrm{i}\phi}$ and correspond (up to scaling), respectively, to the numerators of the complex 3-sphere coordinate fields $u$ and $v^*$ from equation (S2.7) [21]. However, the $z$-dependence of the propagating light fields is different from the full $u$, $v$ ($v^*$) fields. This is similar to the design of optical vortex knots [17], whose design is based on more general complex scalar fields (e.g. "Milnor polynomials") which have the same amplitude in the focal plane and same global 3D topology, although the functional form is different.

## Supplementary Note 3: Optical generation and detection details

In this note we give an overview about the digital propagation method that is used to generate transverse planes at different propagation distances around the focus on the camera, the phase and polarization measurement technique and discuss the error rate on basis of their possible origins in the setup.

### 3.1. Measuring the optical skyrmionic hopfion

Digital propagation is an approach that allows for the investigation of different $z$-planes of a 3D structured paraxial light field without mechanically moving the observation plane. The field of interest itself is propagated by varying the Fourier hologram on a SLM [22, 23] as follows. In terms of the Fourier representation, the scalar 3D light field with transverse position $r_\perp$ physically propagating in the $+z$-direction can be represented by $\psi(r_\perp, z) = \mathcal{F}^{-1}(\mathcal{F}(\psi(r_\perp, 0))e^{ik_z z})$ with $\mathcal{F}^{-1}$ denoting the inverse Fourier transform and, in the paraxial approximation, $k_z \approx k - (k_x^2 + k_y^2)/(2k)$.

Since we display a Fourier hologram $\mathcal{F}(\psi(r_\perp, 0))$ per scalar mode on the SLM for generating the skyrmionic hopfion structure in the focus of L3 (see Supplementary Fig. 4b), we are able to digitally propagate the differently polarized scalar fields $\psi(r_\perp, z)$ and, thereby, the total topological texture in the $z$-direction, by multiplying the Fourier holograms at different transverse planes with the according phase factor $e^{ik_z z}$. For our analysis of the skyrmionic hopfion structure, we propagate in the range of $z$ (see Methods) in 100 steps of $100\lambda$ each, for investigating the phase as well as polarization per $z$-plane.

For the polarization analysis per transverse plane, a rotating QWP and fixed horizontally aligned polarizer (Pol) are mounted in front of a camera (see Supplementary Fig. 4c). Depending on the rotation angle $\Theta$ of the QWP, the camera records spatially resolved intensity patterns described by $I(\Theta) = \frac{1}{2}(s_0 + s_1 \cos^2(2\Theta) + s_2 \cos(2\Theta) \sin(2\Theta) + s_3 \sin(2\Theta))$ [24], where $s_0, s_1, s_2, s_3$ are the unnormalized Stokes parameters, $s_0$ is total intensity, $s_j = s_0 S_j$, for $j = 1, 2, 3$. To suppress time-dependent errors, every intensity profile is the arithmetic mean of 10 intensity patterns. The Stokes parameters are measured recording the intensity patterns within the transverse plane of interest for 15 different angles $\Theta$. The resolution of the measurement is given by the camera (1280 x 1024 px), therefore the transverse polarization profile is determined across a single plane [25, 26]. To use the full resolution of the camera, the beam structure is magnified by lens L4 with a factor of 16 onto the camera. The combination of polarimetry across transverse planes with digital propagation (see above) allows the full detection of the 3D polarization texture in the focal volume.

In order to retrieve not only the polarization but also the phase information from the realised 3D vectorial light field, in addition we implement digital holographic phase measurements [27]. This method allows for the determination of the transverse phase structure in a fixed $z$-plane of one orthogonal component (scalar field) of the beam relative to a reference beam. For phase metrology, we coherently interfere the scalar beam (signal beam), $\psi(r_\perp, z)$, off-axis with a reference beam, $\psi_{\text{ref}}(r_\perp) = |\psi_{\text{ref}}|e^{ik_R R}$ (plane wave). Both beams are of the same circular polarization. The reference beam illuminates the camera at an angle $\Phi$ to the signal beam (see Supplementary Fig. 4d), generating an interference pattern whose intensity is $I(r_\perp) = |\psi(r_\perp) + \psi_{\text{ref}}(r_\perp)|^2$. The mixed term $|\psi||\psi_{\text{ref}}|e^{i(\phi - k_R R)}$ embedded in this interference pattern contains the desired phase information $\chi(r_\perp)$. This term is extracted by a suited low-pass filter located at $k_R$, the $k$-vector of the reference beam. By centering and performing an inverse Fourier transform of this cropped pattern we receive the pure phase profile of the signal beam. Similar to polarimetry, the beam is magnified and the intensity measurements are averaged.

### 3.2. Errors in measurements

In general, various systematic errors might occur within the experiment. For instance, these may include time dependent fluctuations within the system for the generation and measurement of the skyrmionic hopfion structure, and constant inaccuracies of optical components. We observe imperfections of the beam's intensity profile even before the beam is reflected on the SLM. These imperfections are from the laser source itself and the optical components that guide the beam to the SLM. To address this issue, a beam cleaning is installed immediately before the SLM to provide the best possible profile. During its way through the setup, the beam profile is affected by polarizing optics (wave plates and polarizer) and imperfections in the SLM. We minimize these systematic errors by precise adjustment, implementation of high-quality optical components, as well as phase profile corrections [28].

Within the measurement procedure of our VFFR method, minor deviation per pixel for polarization and phase measurements can be observed. The average error per pixel within the transverse plane is experimentally determined to be $< 4\%$ and $< 2\%$ for polarization and phase measurements, respectively. Since we use the arithmetic mean of ten measurements for each picture, this errors represent the standard deviation within ten measurements at the same $(r_\perp, z)$-position within the beam. Note that this average error is calculated from the errors per pixel, which vary spatially. As representative examples, Supplementary Fig. 7 shows a density plot at the focal plane ($z = 0$) of the deviation of the polarization component $S_3$ of the skyrmionic hopfion structure (a) and the phase $\chi_L$ (b).

Errors particularly affect the polarization measurements around the nodal line of $E_R$ (i.e. the LH C line loop). The superposition of $E_R$ and $E_L$ in this area is very sensitive to small fluctuations in the intensity and phase of one of the beams. Slight fluctuations cause rotations of the polarization profile. However, observed deviations in polarization do not hinder the formation of the topology. On the other hand, the phase structure reveals a larger error close to the optical axis, around the phase singularity in $E_L$ (i.e. the axial RH C line). This effect is due to the higher phase gradient in this area, which causes small spatial vibrations in the experimental setup to have a significant effect on the phase variations between repetitive measurements. In spite of this, the phase singularity propagates close to the optical axis, as visible in Supplementary Fig. 6. Thus, it is in a good agreement with the theoretical prediction.

Any errors here are inherited in the determination of the experimental Skyrme number (see Methods). Despite visual deviations between the experimental and the theoretical structure in light and described errors, we were able to realise an optical skyrmionic hopfion with a small deviation of Skyrme number to the theoretically determined value (see Sec. 5.2).

## Supplementary Note 4: Skyrme density, helicity and optical currents

Here we describe in more detail the Skyrme density, the continuous topological charge density which represents the covering of the optical hypersphere by the position-dependent phase and polarization in the optical field volume. Starting with its mathematical definition as the jacobian of the transformation between 3D configuration space and the optical hypersphere, we then put it into a more physical context related to singular optics and topological helicity. This extends to a discussion of the chosen volume over which the Skyrme number is calculated, and then further theoretical features of the model.

### 4.1. Skyrme density as the jacobian of the mapping from real space to optical hypersphere

We describe the basic derivation of the Skyrme density as the jacobian of the topological $\Pi_3$ mapping from the experimental real space volume to the optical hypersphere. This is motivated by the analogous 2-dimensional mapping to the Poincaré sphere.

#### 4.1.1. Baby polarization skyrmion density and covering the Poincaré sphere

The Poincaré sphere is parametrised by azimuth angle $\alpha$, $-\pi < \alpha \leq \pi$, and polar angle $\beta$, $0 \leq \beta \leq \pi$ (defined in equations (S1.10) and (S1.11)). Its area element is, as usual, $\sin\beta \, d\alpha \, d\beta = d\alpha \, d(\cos\beta)$.

As a 2-surface, it has a standard "east-north" or "south-east" orientation. This corresponds to the sense of increase of azimuth $\alpha$ in the northern hemisphere, and its opposite in the southern, as discussed in Sec. 1.2. This means that a mapping from the $x, y$-plane to the Poincaré sphere acquires a jacobian factor, which we call the *2D Skyrme density* $\Sigma^{(2D)}$,

$$\Sigma^{(2D)} \, dx \, dy = \frac{1}{4\pi} \, d\alpha \, d(\cos\beta), \tag{S4.1}$$

where a normalizing factor of the area of the 2-sphere has been included on the right-hand side. Therefore, including the orientation, $\Sigma^{(2D)} = \partial_x\alpha\partial_y(\cos\beta) - \partial_y\alpha\partial_x(\cos\beta) = \hat{z} \cdot \nabla(\alpha) \times \nabla(\cos\beta)$. From the second equality, we define the *baby* *skyrmion density vector*

$$\Sigma^{b} = \nabla\alpha \times \nabla(\cos\beta). \tag{S4.2}$$

Thus $\Sigma^{(2D)} = \Sigma^{b}_{z}$, and in general the density of covering the Poincaré sphere in *any* plane in 3D, with normal vector $\hat{\boldsymbol{u}}$, can be found using $\hat{\boldsymbol{u}} \cdot \Sigma^{b}$. This can be compared to the continuous 2D topological charge density in full Poincaré-type beams [29–31] and baby skyrmions in other systems [32–38].

Full Poincaré beams are propagating beams with position-dependent polarization such that the integral of $\Sigma^{b}_{z}$ in each transverse plane integrates to an integer; this is guaranteed when, as $x^2 + y^2 \to \infty$, the polarization state tends to a constant independent of direction. This effectively allows "point compactification" of the polarization states in the plane, and thus a topological mapping from 2D real plane to 2-sphere [39].

### 4.1.2. Skyrme density and covering the optical hypersphere

A skyrmionic hopfion is a mapping from 3D real space to the optical hypersphere, where the position-dependent optical polarization and phase parameters wrap around the hypersphere. We can find the corresponding 3D Skyrme density by analogy with the 2D case.

We therefore are considering a mapping from 3D real space with cartesian coordinates $x, y, z$ to the optical hypersphere with polarization parameters $\alpha, \beta$ and phase parameter $\gamma$ (defined in equation (S1.13)). We discussed previously, in Sec. 1.2, that the 3-sphere of unit radius has total volume $2\pi^2$, and as derived in equation (S1.24), the normalized volume element of the 3D hypersphere is $(2\pi^2)^{-1} d^3\Omega = (16\pi^2)^{-1} d\alpha\, d(\cos\beta)\, d\gamma$, with positive volume orientation given by $(\alpha, \cos\beta, \gamma)$. Therefore, by analogy with equation (S4.1), the 3D *Skyrme density* $\Sigma$ is the jacobian

$$\Sigma\, dx\, dy\, dz = \frac{1}{16\pi^2}\, d\alpha\, d(\cos\beta)\, d\gamma, \tag{S4.3}$$

where again the right hand side is normalized so the integral over the whole 3-sphere gives unity.

It would also be possible to use 3-sphere polar angles $\varphi, \theta, \omega$ as in equation (S1.18) to define the Skyrme density, but the polarization parameters $\alpha, \beta, \gamma$ are physically more natural to describe the optical state. These latter parameters are those of the Hopf fibration of the 3-sphere: a small interval $\delta(\cos\beta) = \delta S_3$ (corresponding to a thin 'vertical' slice of the Poincaré sphere) corresponds to a thickened torus in the 3-sphere, on which curves of fixed $\gamma$ and fixed $\alpha$ wind in opposite directions as Hopf circles.

In a skyrmionic hopfion, the $\alpha, \beta, \gamma$ parameters wind around their angles in a sense preserving overall positive orientation, guaranteeing the integral of $\Sigma$ over 3D real space entirely covers the optical hypersphere an integer number of times, defining the Skyrme number.

This full 3D skyrmion is similar to topological textures in other systems. A Shankar monopole, for instance, is a texture in fields characterised by a 3D orthogonal frame, i.e. an orthogonal $3 \times 3$ matrix [40–42]. The space of such matrices, the Lie group SO(3), is doubly covered by the 3-sphere; in terms of our parameters, $\gamma$ closes after a $2\pi$ cycle, not $4\pi$.

Other textures of interest are pure hopfions, characterised by mappings from 3D real space to the 2D sphere (such as fields of unit vectors representing magnets, or directors representing liquid crystal orientations). Their relation with our optical Skyrme density is discussed in the next section.

### 4.2. Polarization hopfions and optical currents

A characteristic feature of any topological texture of closed filaments filling 3D space is that they are mutually linked – inherited by the map between real space and the 3-sphere parametrised by the Hopf fibration which has this property. For polarization textures we can consider this in the context of topological features usually considered in singular optics such as optical vortices and C lines.

Optical vortices are zeros of a complex scalar field $\psi$ representing a polarized scalar component. Mathematically, the points $\boldsymbol{r}$ satisfying $\psi(\boldsymbol{r}) = 0$, that is, the *preimages of* 0. These correspond to the set of vortex lines, which can be infinite lines, loops, or knots [16]. In fact, any complex constant $\psi_0$ has a preimage defined by $\psi(\boldsymbol{r}) = \psi_0$, and, just like vortex lines, these are typically filaments in 3D.

A similar argument holds for polarization filaments, where now the complex scalar is $\Psi = E_{\rm L}/E_{\rm R} = \psi_{\rm L}/\psi_{\rm R}$ as defined in equation (S1.14), whose argument is $\alpha$ and modulus is a monotonic function of $\beta$ and $S_3$. However, on LH C lines where $E_{\rm R} = 0$, $\Psi$ may take the value $\infty$. (This notion may be made more mathematically rigorous with the notion of sections of complex bundles, which we avoid to remain close to standard terminology in optics.) As discussed in Supplementary Note 1, this means the parameter space of the map corresponds topologically to a sphere, allowing Hopf textures and, with phase, 3D skyrmions.

The tangent to the filament at each point is given by the *vorticity field*

$$\boldsymbol{\omega} = \frac{-\mathrm{i}}{2} \frac{\nabla \Psi^* \times \nabla \Psi}{(1 + \Psi^* \Psi)^2}. \tag{S4.4}$$

This is similar to the vorticity field $\boldsymbol{\omega}^{sc} \equiv -\mathrm{i}\, \nabla \psi^* \times \nabla \psi$ for a normal complex scalar field $\psi$ [17], but here with an extra denominator regularising the divergence in the neighbourhood of $\Psi = \infty$. Such a field is the curl of an optical current associated with the scalar $\Psi$,

$$\boldsymbol{J} = \frac{-\mathrm{i}}{4\pi} \frac{\Psi^* \nabla \Psi - \Psi \nabla \Psi^*}{1 + \Psi^* \Psi}, \tag{S4.5}$$

and evidently $\boldsymbol{\omega} = \nabla \times \boldsymbol{J}$. Again, apart from the regularising denominator, this is the standard current associated with a complex scalar field, $\boldsymbol{J}^{sc} = \mathrm{Im}\, \psi^* \nabla \psi$.

In the theory of knotted fields, the linking of field lines of divergence-free vector fields are often of interest; we write such a field as a "magnetic field" $\boldsymbol{B}$ (although we are interested here in $\boldsymbol{\omega}$). The continuous linking density of the field lines of $\boldsymbol{B}$ is given by the *helicity density* $\boldsymbol{B} \cdot \boldsymbol{A} = \boldsymbol{A} \cdot \nabla \times \boldsymbol{A}$ for "vector potential" $\boldsymbol{A}$ [38, 43–46]. The integral over all 3D real space gives the total helicity of the field, for a magnetic field related to duality symmetry, and for polarization textures with tangents specified by $\boldsymbol{\omega}$, related to the global linking of polarization filaments.

However, it is clear from equation (S4.5) that the corresponding polarization helicity density $\boldsymbol{J} \cdot \boldsymbol{\omega} = 0$ everywhere except where $\Psi = \infty$, which has not been suitably regularised. An approach to this was considered in [47] (following from similar calculations in [48]), where the singularity on the LH C line in $\boldsymbol{J}$ is resolved by a singular gauge transformation

$$\boldsymbol{J} \rightarrow \boldsymbol{J} + \frac{-\mathrm{i}}{4\pi} \nabla \log \frac{E_{\mathrm{R}}}{E_{\mathrm{R}}^*} \equiv \widetilde{\boldsymbol{J}}. \tag{S4.6}$$

This new current $\widetilde{\boldsymbol{J}}$ again satisfies $\nabla \times \widetilde{\boldsymbol{J}} = \boldsymbol{\omega}$, but now the helicity is nonzero. Using equation (S4.6), the new current is

$$\widetilde{\boldsymbol{J}} = \frac{1}{2\pi} \mathrm{Im}(E_{\mathrm{R}}^* \nabla E_{\mathrm{R}} + E_{\mathrm{L}}^* \nabla E_{\mathrm{L}}) = \frac{1}{2\pi} \frac{\mathrm{Im}(\psi_{\mathrm{R}}^* \nabla \psi_{\mathrm{R}} + \psi_{\mathrm{L}}^* \nabla \psi_{\mathrm{L}})}{(|\psi_{\mathrm{R}}^2| + |\psi_{\mathrm{L}}^2|)}, \tag{S4.7}$$

that is, the regularised current $\widetilde{\boldsymbol{J}} = \boldsymbol{J}_{\mathrm{o}}$, the *orbital current* associated with the field; the numerator is the sum of the two scalar currents associated with the two field components, appropriately normalized. Added to the spin current (which does not play a significant role in the present situation), this gives the Poynting vector of the paraxial optical field [49].

How is this related to the Skyrme density of the previous part? In terms of the angle parameters $\alpha, \beta, \gamma$, it is easy to see that $\mathrm{Im}\, E_{\mathrm{R}}^* \nabla E_{\mathrm{R}} = \frac{1}{4}(1 + \cos\beta)(\nabla\gamma - \nabla\alpha)$ and $\mathrm{Im}\, E_{\mathrm{L}}^* \nabla E_{\mathrm{L}} = \frac{1}{4}(1 - \cos\beta)(\nabla\gamma + \nabla\alpha)$, implying that

$$\boldsymbol{J}_{\mathrm{o}} = \frac{1}{4\pi}(\nabla\gamma - \cos\beta\nabla\alpha), \tag{S4.8}$$

and

$$\boldsymbol{\omega} = \nabla \times \boldsymbol{J}_{\mathrm{o}} = \frac{1}{4\pi} \nabla\alpha \times \nabla(\cos\beta) = \Sigma^{\mathrm{b}}, \tag{S4.9}$$

that is, the vorticity of the polarization field is the baby skyrmion density vector (see equation (S4.2)). This is elegant but is not surprising, as the vorticity vector arises as the jacobian between real space and field space when counting (discrete) topological charge densities in singular optics [3, 50]. As discussed in the main text, the topological role of helicity in flows has a similar expression in terms of an appropriate velocity field [32, 45, 47, 51–53].

We can summarise the preceding discussion in terms of Skyrme density as follows. The directions of the polarization filaments are given by a vector field determined by $\boldsymbol{\omega} = \Sigma^{\mathrm{b}}$. The natural vector potential associated with this divergence-free field is singular at the singularities of $\alpha$ (i.e. the C lines), but when regularised by adding the phase gradient $\nabla\gamma$, the polarization helicity density reduces to the 3D Skyrme density, as indeed it must:

$$\Sigma = \nabla\gamma \cdot \Sigma^{\mathrm{b}} = \boldsymbol{J}_{\mathrm{o}} \cdot \boldsymbol{\omega}, \tag{S4.10}$$

also equivalent to equation (3) in the main text.

The preceding discussion also illustrates that further gauge transformations—that is, adding nonsingular scalar fields to $\gamma$—does not change the overall Skyrme number. In the study of pure hopfions (i.e. maps from 3D real space to a 2-sphere), whose filament directions are given by $\boldsymbol{B}$ but there is no natural phase, it is usually necessary to construct a 'vector potential' to get the correct helicity density [44, 46].

Similarly, if we choose to focus purely on optical polarization hopfions, we can choose an appropriate phase. The associated helicity equation (S4.10) integrated over all 3D space characterises the interlinking of polarization filaments, and globally the number of times the 3D field wraps around the Poincaré sphere. Any convenient choice of phase function (which is appropriately singular on $\mathbb{C}$ lines) suffices for this choice. Several choices are discussed in the following section, which applies the preceding analysis to our model field, as introduced in the Methods and Supplementary Note 2.

We also remark on textures for a normal (non-infinity) complex scalar field $\psi$. Assuming that $\psi$, as a regular complex-valued function, is nowhere infinite (not even at spatial infinity) means the total set of complex values the field takes is of finite range for a finite set, and hence the texture of its filaments (labelled by $\psi_0$) cannot be topological in spite of some filaments being knotted and linked. If we choose to regarding the vortex lines of $\psi$ (i.e. preimages of zero) being special, we can choose the phase gradient $\nabla \arg \psi$, analogous to the superfluid velocity, as the appropriate current vector. The curl of $\nabla \arg \psi$ vanishes everywhere apart from $\delta$-concentrations on the vortex line; careful regularisation gives a weighting to the screw dislocation nature of the wave vortex [50]. As the global helicity of the field must be zero, the phase around a vortex loop must unwind (i.e. "self-link") whenever it is threaded (i.e. "linked") with another vortex line [54].

### 4.3. Integrating the Skyrme and Hopf densities of the model of skyrmionic hopfion

The topology of the optical skyrmionic hopfion resides in the map from 3D configuration space to the optical hypersphere, parametrised by $\alpha, \beta, \gamma$ by the fields (S2.9), (S2.11) with the numerical values of the coefficients given in the Methods. As discussed in the main text in the context of the experiment, the polarization of these fields consists of a texture of filaments of constant polarization resembling the Hopf fibration (main text Fig. 3), and phase varying along them establishing a $\Pi_3$ mapping between the two 3D spaces. The Skyrme density is given by $\Sigma = \frac{1}{16\pi^2} \nabla \gamma \cdot (\nabla \alpha \times \nabla (\cos \beta))$.

As described in the Methods, the coefficients of this model field have been chosen optimally to concentrate the intensity and Skyrme density. Supplementary Fig. 8a shows a cross-section of the Skyrme density for this field (equivalent to the 3D plot in the inset to Fig. 4c in the main text). The Skyrme density for the model field is strongly concentrated around the $\mathbb{C}$ line ring, with most of the contours shown in the plot being several orders of magnitude smaller than the maximum density.

Provided the neighbourhood of the vortex ring is included in the integration volume, we expect the integrated Skyrme density to be close to 1. Supplementary Fig. 8b demonstrates the result of this integral as a function of the transverse ($L_\perp$) and longitudinal ($L_\parallel$) dimensions of the integration volume given by $|x| \leq L_\perp$, $|y| \leq L_\perp$, $|z| \leq L_\parallel$. Most of the area shows 99% of the 3-sphere volume is filled. This justifies our claim that the optical skyrmionic hopfion is particle-like – the integration of the topological charge density is independent of the size of the sufficiently large integration box, and is very close to unity.

The evaluation of the measured Skyrme number from the experimental data is described in Methods and Supplementary Note 5. As explained there, rather than construct a smooth interpolation of $\Sigma$ from the data then integrate numerically, we calculate the fraction of the 3-sphere filled by a piecewise linear map constructed directly from the data. Even with high-performance computers this approach requires computational effort. Justified by the particle-like nature of the skyrmionic hopfion, we limit the calculation of the experimental Skyrme number to the cuboid of transverse size $L_\perp = 1.84w = 99.4\,\mu\text{m}$ and longitudinal size $L_\parallel = 0.768z_R = 26.6\,\text{mm}$ indicated in Supplementary Fig. 8, a and b, in cyan. The numerical integral of the model field in this restricted cuboid is 0.9966, i.e. 99.7% of the 3-sphere volume. For reference, the total measured volume (see Methods) is the cuboid given by $x_{\max} = 3.13w = 170\,\mu\text{m}$, $y_{\max} = 3.91w = 212\,\mu\text{m}$ and $z_{\max} = 0.768z_R = 26.6\text{mm}$. The numerical integral of $\Sigma$ for the model field in the measured volume is 0.997 and is also indicated in Supplementary Fig. 8b in cyan. Thus, to 3 significant figures, the Skyrme number in the reduced volume is no different from the integrated density in the total measured volume.

We note that there is a contour in Supplementary Fig. 8b for $L_\perp > 2w$, $L_\parallel > 1.5z_R$ where the integrated Skyrme volume equals 1. This due to the properties of gaussian beams used to realise the field, which is described, with a mathematical analysis of the asymptotic value of the field for large $R$ and $z$, in the next subsection. However, for practical purposes it is challenging to measure the field's polarization and phase so far from the focal point, as the intensity is small and subject to experimental imperfections in this region.

### 4.4. Theoretical considerations of model skyrmionic hopfion

We now consider some of the topological details associated with our model field (S2.9), (S2.11) in more detail. For convenience, we use dimensionless cylindrical coordinates $\overline{R} = R/w$, $\overline{z} = z/z_R$, but keeping the coefficients $a, b, c$. To simplify the expressions, we write $1 + \overline{z}^2 = W$.

The Skyrme density $\Sigma = \frac{1}{16\pi^2} \nabla\gamma \cdot (\nabla\alpha \times \nabla(\cos\beta))$ for the superposition of LG modes given by equations (S2.9) and (S2.11), in terms of these dimensionless coordinates, is

$$\Sigma = \Big( c^2 (a^2(2 - \overline{R}^2)\overline{z}W^3 + b^2(\overline{R}^6\overline{z}W + 16\overline{z}^2\overline{z}W^2 - 2\overline{R}^4(3 - \overline{z}^2) + 4\overline{R}^2(1 - 3\overline{z}^2 - 5\overline{z}^4 - \overline{z}^6))$$
$$-4ab(2\overline{R}^4\overline{z}^2 - \overline{z}W^2(1 - 3\overline{z}^2) + \overline{R}^2(1 - 3\overline{z}^2 - 5\overline{z}^4 - \overline{z}^6)))\Big)$$
$$/\pi^2\overline{z}W(4c^2\overline{R}^2\overline{z}W + a^2\overline{z}W^2 + b^2(4\overline{z}^2 + (\overline{R}^2 - 2\overline{z}^2)^2) - 2ab(\overline{R}^2(1 - \overline{z}^2) + 2\overline{z}^2\overline{z}W))^2. \qquad \text{(S4.11)}$$

For $\overline{R} \gg 1$, we have that $\Sigma \sim c^2/\pi^2 b^2 \overline{R}^2 + O(1/R^4)$. Therefore the integral over infinite radius diverges logarithmically. We will consider the meaning of this in the following.

#### 4.4.1. Polarization parameters

From equation (S1.10), the gradient $\nabla\alpha$ for the model field in cylindrical coordinates is

$$\nabla\alpha = -\frac{\widehat{\boldsymbol{\phi}}}{\overline{R}} + \frac{\left[ 4b(b-a)\overline{R}\overline{z}(1 + \overline{z}^2)\widehat{\boldsymbol{R}} + 2ab(1 + 2\overline{z}^2(1 + \overline{R}^2) + \overline{z}^4 - a^2(1 + \overline{z}^2)^2 - b^2\overline{R}^2(2 + 6\overline{z}^2 - \overline{R}^2)\widehat{\boldsymbol{z}} \right]}{(1 + \overline{z}^2)((a - b\overline{R}^2)^2 + 2\overline{z}^2(a^2 - ab(2 - \overline{R}^2) + 2b^2(1 - \overline{R}^2)) + (a - 2b)^2\overline{z}^4)}, \quad \text{(S4.12)}$$

which is singular along and circulates around the C lines: along the axis $\overline{R} = 0$ and the circle $(\overline{R}, \overline{Z}) = (\sqrt{a/b}, 0)$.

From (S1.11), $\cos\beta = S_3$, which is given by

$$S_3 = 1 - \frac{8c\overline{R}^2(1 + Z^2)}{4c^2\overline{R}^2\overline{z}W + a^2\overline{z}W^2 + b^2(4\overline{z}^2 + (\overline{R}^2 - 2\overline{z}^2)^2) - 2ab(\overline{R}^2(1 - Z^2) + 2\overline{z}^2\overline{z}W)}. \qquad \text{(S4.13)}$$

Its gradient $\nabla(\cos\beta)$ is

$$\frac{16c^2\overline{R}\left( \overline{z}W(b\overline{R}^4 - \overline{z}W(a^2 + (a - 2b)^2\overline{z}^2))\widehat{\boldsymbol{R}} + \overline{R}\overline{z}(a^2\overline{z}W^2 + 4ab(\overline{R}^2 - \overline{z}W^2) + b^2(4\overline{z}W^2 - \overline{R}^2(4 + \overline{R}^2))\widehat{\boldsymbol{z}} \right)}{\left( 4c^2\overline{R}^2\overline{z}W + a^2\overline{z}W^2 + b^2(4\overline{z}^2 + (\overline{R}^2 - 2\overline{z}^2)^2) - 2ab(\overline{R}^2(1 - \overline{z}^2) + 2\overline{z}^2\overline{z}W) \right)^2}. \qquad \text{(S4.14)}$$

Together, these determine the polarization behavior of the skyrmionic hopfion, and the field's baby skyrmion/full Poincaré beam nature in any plane.

We now consider the Skyrme density and Hopf density (polarization helicity) of the model field for $\psi_R$ and $\psi_L$ given in equation (S2.9) and equation (S2.11) respectively, found as superpositions of LG modes.

#### 4.4.2. 2D baby Skyrme density/full Poincaré beam nature

The model beam's baby Skyrme density (polarization vorticity) vector is, in cylindrical coordinates,

$$\Sigma^b = \frac{4b^2}{\pi((1 + \overline{z}^2)(4b^2\overline{R}^2 + a^2W - 4ab\overline{z}^2) + b^2((\overline{R}^2 - 2\overline{z}^2)^2 + 4\overline{z}^2) - 2ab\overline{R}^2(1 - \overline{z}^2)^2}$$
$$\times \Big[ -b\overline{R}(-1 + \overline{R}^2 - \overline{z}^2)(-a\overline{z}W + b(2 + \overline{R}^2 + 2Z^2))(\widehat{z}\widehat{\boldsymbol{R}} + \widehat{\boldsymbol{\phi}})$$
$$+ \overline{R}(b - a)(a\overline{z}W^2 - b(\overline{R}^2(-3 + \overline{z}^2) + 2\overline{z}W^2))(\widehat{z}\widehat{\boldsymbol{R}} + \widehat{\boldsymbol{\phi}})$$
$$-\overline{z}W(-b^2\overline{R}^4 + \overline{z}W(a^2 + (a - 2b)^2\overline{z}^2)\widehat{\boldsymbol{z}}) \Big]. \qquad \text{(S4.15)}$$

The $z$-component of this vector, $\Sigma_z^b$, gives the density covering the Poincaré sphere in a transverse plane. In the focal plane $\overline{z} = 0$, the RH axial C line corresponds to the origin, and around it is the LH C line of radius $\overline{R} = \sqrt{a/b}$, equal to 0.566 with the numerical coefficients. Since the vortex sign of the LH component is negative, the C point at the origin has a 'star' form [15, 16] − $\alpha$ decreases as $\phi$ increases in a right-handed sense around the beam axis.

With the numerical values of the coefficients, the integral of $\Sigma_z^b$ in the focal plane, over the disk to radius $\sqrt{a/b}$ gives $-1$, equivalent to covering the Poincaré sphere once in a negative sense. This is consistent with the azimuthal sense of $\nabla_\perp \alpha$ being against $\hat{\phi}$ while $\nabla_\perp(\cos\phi) = \nabla_\perp S_3$ points towards the origin within this disk, ensuring $\hat{z} \cdot \nabla_\perp \alpha \times \nabla_\perp(\cos\beta) < 0$. Outside the radius of the C line loop, the sign of $\nabla_\perp S_3$ changes although $\nabla_\perp \alpha$ does not, so $\Sigma_z^b > 0$; since, asymptotically as $\overline{R} \gg 1$, the polarization state is RH, the net baby Skyrme number (degree of the map covering the Poincaré sphere) is zero.

The magnitude of the baby Skyrme density in the focal plane near the circular C line is large; within the circles where $S_3 = \cos\beta = 0$, at $\overline{R} = 0.524, 0.610$ with the numerical coefficients, corresponds to covering the entire lower hemisphere, whereas the entire upper hemisphere is covered inside and outside this annulus.

By topological continuity, the baby Skyrme number in every other transverse plane is also zero (the radial asymptotic state of polarization being RH circular ensures this is always well-defined). As the LH circular polarization is restricted only to the C line loop in the focal plane, the baby Skyrme number must be zero in every transverse plane except the focal plane.

The 3D Skyrme number involves the product of $\Sigma_z^b$ with $\nabla\gamma$. Since the C line is a phase vortex in $\gamma$, $\nabla\gamma$ changes sign on each side of the circular C line, implying the 3D Skyrme density $\Sigma$ maintains the same sign although $\Sigma_z^b$ changes sign.

We briefly remark on the comparison between the 2D baby skyrmions in our transverse beam profiles and the 'Néel type' (radial) and 'Bloch type' (spiral) baby skyrmions studied in magnetic systems [34]. In a magnetic system, (baby) skyrmions are localised, nonlinear excitations of a ground state (corresponding, e.g. to LH polarization) with Skyrme number 1 in our terminology; the 'type' indicates whether azimuth angles correspond to $\phi$ with an offset or not. In our beams, in the focal plane, $\phi = -\alpha$, so there is no spiral-like contribution: the baby skyrmion configuration in the focal plane is purely radial (Néel). However, in other planes, $\alpha$ acquires an offset (which is dependent on radius) due to the phase difference between $E_R$ and $E_L$, meaning in a typical plane it appears Bloch-type. Of course, for polarization, the distinction between radial and spiral is relative; since polarization is a pseudospinor, the relation between polarization azimuth $\alpha$ and real space azimimuth $\phi$ is a matter of convention.

### 4.4.3. Polarization hopfion

We now consider our model field purely as a hopfion, i.e. how the Poincaré sphere is covered by the polarization states in the focal volume. As a map from 3D (real space) to the Poincaré 2-sphere, the preimage of each polarization state is a 1D line − these are the polarization filaments discussed in the main text, generalising C lines for circular polarization.

For an ideal hopfion, the filaments should correspond to a Hopf fibration in real space: loops wound around tori on which $S_3 = \cos\beta$ is constant; these tori enclose the LH circular C line loop. The contours of $S_3$ from equation (S4.13) in a plane of fixed azimuth is shown in Supplementary Fig. 2c. Evidently, $S_3 = -1$ on the LH C line loop (reference circle) with $\overline{R} = r_0 = \sqrt{a/b}$. The ideal hopfion structure would have the contours of constant $S_3$ loops around this point, topologically equivalent to the contours of $|u|$ in Supplementary Fig.2a. With the numerical choice of $a$ and $b$ as discussed in Supplementary Note 2, $S_3$ has saddle points at $(\overline{R}, \overline{z}) = (2\sqrt{b}, \pm\sqrt{a+b})/\sqrt{3b-a} = (1.22, \pm 0.702)$, at $S_3 = 1 - 8c^2/((a-b)^2 + 4c^2) = 0.969$.

At values of $S_3$ larger than this value, $0.969 < S_3 \leq 1$, the toroidal structure beaks down, into topological cylinders extending to $\overline{z} \to \pm\infty$; topologically the hopfion structure breaks down here, although this counts for only 1.5% of the range of $S_3 = \cos\beta$.

As discussed above, the Hopf number of the hopfion can only be calculated with the choice of a suitable phase function $\widetilde{\gamma}$; using the physical phase $\gamma$ gives the skyrmionic hopfion, but the global integral of $\nabla\widetilde{\gamma} \cdot \Sigma^b$ should not change under a gauge transformation $\widetilde{\gamma} \to \widetilde{\gamma}'$. As noted above, the physical phase fails to converge due to the LG beams' wavefront curvature, so we will first consider the physical phase gradient in more detail.

### 4.4.4. Physical phase gradient

The phase parameter $\gamma$ is the sum of phases of the two beam components, from equation (S1.13). Its gradient therefore is a sum of phase gradients from different phase features of the full field, as superposition of LG beams,

$$\nabla\gamma = \boldsymbol{g}^r + \boldsymbol{g}^a + \boldsymbol{g}^G + \boldsymbol{g}^c, \tag{S4.16}$$

where $\boldsymbol{g}^r$ is associated with the phase of the C line ring, $\boldsymbol{g}^a$ with the axial C line, $\boldsymbol{g}^G$ with the Gouy phase, and $\boldsymbol{g}^c$ with the wavefront curvature. The separation into these terms is very natural for our model, consisting of a superposition of three LG beams, with $(\ell, p) = (0, 0), (0, 1)$ and $(−1, 0)$. All of these have the same wavefront curvature factor, and the axial phase is due to the negative vortex in the LH component. Since the beam with $(\ell, p) = (0, 1)$ dominates the superposition for large $R$ and $z$, we choose the Gouy phase of this beam to represent $\boldsymbol{g}^{\mathrm{G}}$; the phase for the ring $\boldsymbol{g}^{\mathrm{r}}$ takes all the other phase contributions, including the contributions from the Laguerre polynomials. Only $\boldsymbol{g}^{\mathrm{r}}$ involves the coefficients of the superposition.

The full Skyrme density $\Sigma = \nabla \gamma \cdot \Sigma^{\mathrm{b}}$ can therefore be considered as the contribution of separate terms, $\Sigma^{\bullet} = \boldsymbol{g}^{\bullet} \cdot \Sigma^{\mathrm{b}}$, for $\bullet = \mathrm{r, a, G, c}$ respectively. The important contributions for the skyrmion topology are those phases corresponding to phase singularities in the physical field (and singularities in $\alpha$), that is, from the C lines. These are therefore $\boldsymbol{g}^{\mathrm{r}}$ and $\boldsymbol{g}^{\mathrm{a}}$.

The various phase gradients are readily found to be

$$\left.\begin{aligned}
\boldsymbol{g}^{\mathrm{r}} &= \tfrac{2(a-b)}{((a-b\overline{R}^2)^2 + 2(a^2 - ab(2-\overline{R}^2) + 2b^2(1-R^2))\overline{z}^2 + (a-2b)^2\overline{z}^4)} \left( 2bR\overline{z}\widehat{\boldsymbol{R}} + (a - bR^2 + (a-2b)\overline{z}^2)\widehat{\boldsymbol{z}} \right), \\
\boldsymbol{g}^{\mathrm{a}} &= -\widehat{\boldsymbol{\phi}}/\overline{R}, \\
\boldsymbol{g}^{\mathrm{G}} &= -5W\widehat{\boldsymbol{z}} \\
\boldsymbol{g}^{\mathrm{c}} &= 2\overline{R}\overline{z}\widehat{\boldsymbol{R}}/W + R^2(1-\overline{z}^2)\widehat{\boldsymbol{z}}/W^2.
\end{aligned}\right\} \tag{S4.17}$$

Evidently, from equation (S4.12), $\nabla\alpha = \boldsymbol{g}^{\mathrm{a}} - \boldsymbol{g}^{\mathrm{r}} + (1+\overline{z}^2)^{-1}\widehat{\boldsymbol{z}}$; the final term accounts for the difference in Gouy phases of the RH and LH beams, and otherwise $\nabla\alpha$ is the difference of the axial and ring phase gradients whilst $\nabla\gamma$ includes their sum, just like how, as described in Supplementary Note 1, $\arg U$ and $\arg V$ are the sum and difference of $\alpha$ and $\gamma$ for the Hopf fibration of the optical hypersphere.

Both $\Sigma^{\mathrm{r}}$ and $\Sigma^{\mathrm{a}}$ must therefore be included in any function determining the Hopf density of the full field. Numerical integration of the cut volumes of $|\overline{x}|, |\overline{y}| \leq 1.84$, $|\overline{z}| \leq 0.768$ gives 0.4964 and 0.4967 for the axial and ring contributions to the Skyrme number, adding to 0.9931.

The other phase gradients $\boldsymbol{g}^{\mathrm{G}}$ and $\boldsymbol{g}^{\mathrm{c}}$ arise from the physics of propagating gaussian beams. The Gouy phase $\boldsymbol{g}^{\mathrm{G}}$ is purely in the $z$-direction, so the corresponding Hopf density $\Sigma^{\mathrm{G}}$ is proportional to the transverse baby Skyrme density (as indeed would be any phase associated with the $e^{ikz}$ propagation); as discussed above, its integral over any transverse plane is zero, meaning the integral of $\Sigma^{\mathrm{G}}$ over all space is zero. Within the investigated volume, $\int_{\mathrm{vol}} \Sigma^{\mathrm{G}}\, \mathrm{d}^3\boldsymbol{r} = 0.0014$, making an almost negligible contribution.

Unlike the other contributions to $\nabla\gamma$, the wavefront curvature phase gradient $\boldsymbol{g}^{\mathrm{c}}$ diverges as $\overline{R} \gg 1$, since the phase curvature does not settle to a constant value for $\overline{R} \gg 1$ when $\overline{z} \neq 0$. In this regime, $\Sigma^{\mathrm{c}} \sim c^2/\pi^2 b^2 \overline{R}^2$, accounting for the total logarithmic divergence in the total $\Sigma$. This contribution accounts for why the integrated Skyrme density exceeds 1 for much larger volumes, as can be seen in Supplementary Fig. 8b when the boundaries $L_\perp$ and $L_\parallel$ are quite large.

Although mathematically a divergence, the actual divergence is very slow, becoming large where the beam intensity is low, and thus hard to measure physically. In fact, the paraxial approximation of the physical beam in terms of the mathematical model breaks down for sufficiently large $\overline{R}$ and $\overline{z}$. Within the cut volume as described above, the wavefront curvature contributes 0.0020 to the Skyrme number; slightly larger than the Gouy phase, but still very small.

The sum of these four contributions gives the value of the Skyrme number numerically integrated over the cut volume, as described above: $0.4964 + 0.4967 + 0.0015 + 0.0020 = 0.9966$.

### 4.4.5. Global Hopf number

From the previous section, it is natural to calculate the global Hopf number as the integral over all space of $\Sigma^{\mathrm{a}} + \Sigma^{\mathrm{r}}$; these include the important topological helicity terms and do not diverge at infinity. With the numerical values of the coefficients, this global Hopf number in fact gives zero; the nontrivial degree of the localised hopfion within the measured volume is, mathematically, unwound within the larger volumes, beyond the saddle of $\cos\beta$ in $\overline{R}, \overline{z}$.

This does not remove the "particle-like" nature of the measured hopfion, although the precise numerical value depends on precisely which phase contributions are included; unlike the Skyrme density, which is physically defined in any volume, the more general Hopf density is only unambiguous within the infinite volume.

Preliminary numerical investigations of beams with different values of $a, b, c$, not constrained by being realised in experiment, indicate that configurations with nonzero global Hopf number are possible. This behavior of the global Hopf number indicates some subtlety in the global topology of such paraxial LG beam superpositions which would merit further investigation.

### 4.4.6. Ideal skyrmion field and higher Skyrme numbers

We conclude this section with a short consideration of the ideal skyrmion based on $u$ and $v$ defined in equation (S2.7),

$$\left.\begin{aligned} E_{\mathrm{R}} &= \left(\frac{R^2+z^2-1+2\,\mathrm{i}\,z}{R^2+z^2+1}\right)^n &&= \mathrm{sech}^n\,\tau\,\mathrm{e}^{\mathrm{i}\,n\sigma}, \\ E_{\mathrm{L}} &= \left(\frac{2R\,\mathrm{e}^{\mathrm{i}\phi}}{R^2+z^2+1}\right)^m &&= \tanh^m\,\tau\,\mathrm{e}^{\mathrm{i}\,m\phi}, \end{aligned}\right\} \tag{S4.18}$$

where $m, n$ are nonnegative integers, the first equalities define the fields in terms of cylindrical coordinates $R, \phi, z$, and the second equalities in terms of toroidal coordinates [11–13], similar to Supplementary Note 1 but here applied to real 3D space. We perform the calculation here in toroidal coordinates, which is very natural for the ideal skyrmion field. This can be compared to the similar calculation in [55], which assumes spherical symmetry.

The parameters of this definition of the field, from equations (S1.10), (S1.11) and (S1.13), are easily expressed in toroidal coordinates,

$$\gamma = m\phi + n\sigma, \qquad \alpha = m\phi - n\sigma, \qquad \cos\beta = \frac{\mathrm{sech}^{2n}\,\tau - \tanh^{2n}\,\tau}{\mathrm{sech}^{2n}\,\tau + \tanh^{2n}\,\tau}. \tag{S4.19}$$

Continuing in toroidal coordinates, the gradients are

$$\left.\begin{aligned} \nabla\gamma &= (\cosh\tau - \cos\sigma)(m\hat{\boldsymbol{\phi}}/\sinh\tau + n\hat{\boldsymbol{\sigma}}), \\ \nabla\alpha &= (\cosh\tau - \cos\sigma)(m\hat{\boldsymbol{\phi}}/\sinh\tau - n\hat{\boldsymbol{\sigma}}), \\ \nabla(\cos\beta) &= -4(\cosh\tau - \cos\sigma)\,\mathrm{csch}^n\,\tau\,\mathrm{sech}^{2m+1}\,\tau\,\tanh^{2n}\,\tau\,\hat{\boldsymbol{\tau}}/(\mathrm{sech}^{2n}\,\tau + \tanh^{2n}\,\tau)^2. \end{aligned}\right\} \tag{S4.20}$$

The volume element for toroidal coordinates is $\sinh\tau\,\mathrm{d}\phi\,\mathrm{d}\sigma\,\mathrm{d}\tau/(\cosh\tau - \cos\sigma)^3$, so the Skyrme density $\Sigma = \nabla\gamma \cdot \nabla\alpha \times \nabla(\cos\beta)$, and dividing by the 3-sphere volume $16\pi^2$, gives the total integrated Skyrme number

$$\int_0^\infty \int_{-\pi}^{\pi} \int_{-\pi}^{\pi} \frac{\Sigma\sinh\tau}{16\pi^2(\cosh\tau - \cos\sigma)^3}\,\mathrm{d}\phi\,\mathrm{d}\sigma\,\mathrm{d}\tau = mn \int_0^\infty \frac{\mathrm{d}}{\mathrm{d}\tau}\left[\frac{1}{1 + \cosh^{2n}\,\tau\,\tanh^{2m}\,\tau}\right]\mathrm{d}\tau = -mn, \tag{S4.21}$$

where, since the net integrand is dependent only on $\tau$, the angle integrals are trivial. The net integral is identified as a pure derivative, so the overall $\tau$ integral is also straightforward, leaving a net result $mn$ from the phase windings in equation (S4.18). Choosing the conjugate of either of the fields in the definition (S4.18) gives a positive Skyrme density, as in previous discussion where $m = n = 1$ and a negative axial vortex was chosen.

This is instructive for designing optical skyrmionic hopfions of higher Skyrme number, e.g. using superpositions of LG beams. The strategy is similar to that employed by [56], who designed scalar fields with vortex loops threaded by vortices on the axis. Increasing the axial degree $m$ is easy, by replacing the LH component with a different azimuthal quantum number $\ell = m$. However, as described in [56], paraxial beams do not allow transverse vortex loops with $|n| > 1$, Realising a higher-order skyrmion with effective $|n| > 1$ is possible with multiple vortex loops with the same sense of vorticity, and indeed, these can be woven into various kinds of link and knot as constructed in [47, 48].

### Supplementary Note 5: Details of numerical methods

Given the the map from real space to the optical hypersphere described in the numerical calculation of the experimental Skyrme Number (see Methods), we can visualise the optical skyrmionic hopfion and its Skyrme density. We imagine each measured point in real space to correspond to the vertices of a cubic lattice, which is carried to an irregular lattice with the same topology in the optical hypersphere. In regions of the beam where the Skyrme density is high, such as close to the circular C line, the corresponding spherical tetrahedra (and cubes comprised of them) are large, and where the Skyrme density is small so are the spherical tetrahedra. Explicative examples are the grey cubes in Fig. 4 of the main text: in real space (Fig. 4a) the cubes have the same size, whereas in the optical hypersphere, (Fig. 4b), the cube enclosing the C line loop (in black) is much larger than the other cube. Supplementary Fig. 9, a and b, show the analogous mapping for the ideal skyrmion field. Main text Fig. 4c shows all of the sampled cubes in real space, whose colour represents $\mathrm{Vol}(C)$: the transparency goes to 0 for very small volumes, becomes increasingly orange for positive volumes and increasingly blue for negative volumes. Our algorithm is based on expressing the measured data as the vertices of tetrahedral meshes (i.e. 3D cell complexes) in real space and the optical hypersphere. The calculations of the spherical tetrahedra volumes in the 3-sphere are described, with a final section describing how this is found from the image of the boundary of the measured volume.

### 5.1. The volume of a spherical tetrahedron

The 3-sphere is embedded in four-dimensional Euclidean space. Four unit vectors, $\vec{n}_\ell$ with $\ell = \{a, b, c, d\}$, point to the four vertices of a spherical tetrahedron $\mathcal{T}$. The four faces of this spherical tetrahedron are spherical triangles, pairs of which meet along edges at the dihedral angles $0 < \varphi_j < \pi$ for $j = 1, 2, \ldots, 6$. The formula for $\mathrm{Vol}(\mathcal{T})$ [57], 3D volume of the tetrahedron in the hypersphere, is in terms of the dihedral angles.

Given $a_j = \exp(i\varphi_j)$ and the dilogarithm function $\mathrm{Li}_2(z)$, the volume of the spherical tetrahedron is, from Murakami's formula [57],

$$\mathrm{Vol}(\mathcal{T}) = -\,\mathrm{Re}(\mathsf{L}(a_1, a_2, a_3, a_4, a_5, a_6, z_0)) + \pi \left( \arg(-q_2) + \frac{1}{2} \sum_{j=1}^{6} \varphi_j \right) - \frac{3}{2}\pi^2 \bmod 2\pi^2, \qquad (S5.1)$$

where

$$\mathsf{L}(a_1, a_2, a_3, a_4, a_5, a_6, z) = \frac{1}{2} \Big( \mathrm{Li}_2(z) + \mathrm{Li}_2(a_1^{-1}a_2^{-1}a_4^{-1}a_5^{-1}z) + \mathrm{Li}_2(a_1^{-1}a_3^{-1}a_4^{-1}a_6^{-1}z) + \mathrm{Li}_2(a_2^{-1}a_3^{-1}a_5^{-1}a_6^{-1}z) $$
$$ -\mathrm{Li}_2(-a_1^{-1}a_2^{-1}a_3^{-1}z) - \mathrm{Li}_2(-a_1^{-1}a_5^{-1}a_6^{-1}z) - \mathrm{Li}_2(-a_2^{-1}a_4^{-1}a_6^{-1}z) - \mathrm{Li}_2(-a_3^{-1}a_4^{-1}a_5^{-1}z) + \sum_{j=1}^{3} \log a_j \log a_{j+3} \Big) \quad (S5.2)$$

and

$$z_0 = -q_1 + \sqrt{q_1^2 - 4q_0 q_2}/(2q_2),$$
$$q_0 = a_1 a_4 + a_2 a_5 + a_3 a_6 + a_1 a_2 a_6 + a_1 a_3 a_5 + a_2 a_3 a_4 + a_4 a_5 a_6 + a_1 a_2 a_3 a_4 a_5 a_6,$$
$$q_1 = -(a_1 - a_1^{-1})(a_4 - a_4^{-1}) - (a_2 - a_2^{-1})(a_5 - a_5^{-1}) - (a_3 - a_3^{-1})(a_6 - a_6^{-1}),$$
$$q_2 = a_1^{-1}a_4^{-1} + a_2^{-1}a_5^{-1} + a_3^{-1}a_6^{-1} + a_1^{-1}a_2^{-1}a_6^{-1} + a_1^{-1}a_3^{-1}a_5^{-1} + a_2^{-1}a_3^{-1}a_4^{-1} + a_4^{-1}a_5^{-1}a_6^{-1} + a_1^{-1}a_2^{-1}a_3^{-1}a_4^{-1}a_5^{-1}a_6^{-1}. \qquad (S5.3)$$

The dihedral angles are found from the tetrahedron's vertices using hyperspherical trigonometry as follows [58]. For each triple of hyperplanes with normals $\vec{n}_i, \vec{n}_j, \vec{n}_k$, define the vector $\vec{u}_{ijk}$ orthogonal to each hyperplane, by

$$\vec{u}_{ijk} = (\vec{n}_i \times \vec{n}_j \times \vec{n}_k)/|\vec{n}_i \times \vec{n}_j \times \vec{n}_k|.$$

Here, the ternary vector product in four dimensions, $(\vec{a} \times \vec{b} \times \vec{c})$, can be defined as $(\vec{a} \times \vec{b} \times \vec{c}) \cdot \vec{e}_i = \det(\vec{a}, \vec{b}, \vec{c}, \vec{e}_i) = \varepsilon_{ijkl}a_j b_k c_l$, where $\vec{e}_i$ are the unit vectors in four dimensions and $\varepsilon_{ijkl}$ is the Levi-Civita symbol [58]. This is related to the 4D volume form considered in Supplementary Note 1.

Each dihedral angle $\varphi_{jk}$ satisfies $\cos\varphi_{jk} = \vec{u}_{ijk} \cdot \vec{u}_{jkl}$. The two indices of the dihedral angles are simplified to a unique index $j = (1, 2, \ldots, 6) = (da, db, cd, bc, ac, cb)$, which, when substituted into the formula (S5.1), gives the volume of a single tetrahedron.

### 5.2. Visualising the covered volume of optical hypersphere

We previously showed in Supplementary Note 4 that the integrated Skyrme density for the same measured volume corresponds to is 0.997 of the hypersphere volume. This difference can be accounted for from experimental imperfections and the Skyrme number. Topologically, the difference must be related to the images of the measured volume boundary in the hypersphere.

The optical states on the boundary of the measured volume are to some extent arbitrary. The optical state $(E_R, E_L) = (-1, 0)$ (i.e. $\omega = \pi$) occurs at the focal point, and as $R \to \infty$ in the focal plane $z = 0$, $(E_R, E_L) = (+1, 0)$ (i.e. $\omega = 0$). The $\mathrm{LG}_{0,1}$ contribution to $E_R$ dominates the field for large $R \gg w$, so for an appropriately large measurement volume (although not too large, for reasons discussed in Sec. 4.4), the boundary of the measurement volume consists of optical states close, in the optical hypersphere, to the state $(1, 0)$, which is antipodal to the state at the focal point.

The image of the measurement boundary separates the optical hypersphere into two disjoint volumes: that covered by the measured optical states (covering 94 % experimentally), and a neighbourhood of the state $(1, 0)$, covering approximately 6 % of the hypersphere. We chose to represent the optical hypersphere, in main text Figs. 3 and 4, using the volume-preserving projection with the optical state at the focal point, $(-1, 0)$. The antipodal state $(-1, 0)$ therefore corresponds to the sphere boundary of the solid ball. The image of the measured boundary, being a neighbourhood of this point, therefore would appear to be a sphere close to the boundary sphere in this projection. The boundary of the measured volume, that is, the image of the six faces of the measured cubic volume, are plotted in the described projection of the optical hypersphere in Supplementary Fig. 10a. Clearly this is a surface that almost coincides with the boundary of the projection at $(1, 0)$. The corresponding plot for the theoretical model field is shown in the inset.

In this projection it is hard to see that the volume of this neighbourhood complementing the filled volume is small. For this reason, in Supplementary Fig. 10b, we plot the optical hypersphere in a different projection, with a point on the circular C line at the center (and the opposite point on the C line loop to the boundary of the ball). The RH states on the axis in real space become a circle in this projection, and it is clear here that the volume of the optical hypersphere not covered is small (consistent with the theoretical model, inset to Supplementary Fig. 10b).

## Supplementary References

[1] Born, M. and Wolf, E. *Principles of Optics*. Cambridge University Press, 7th edition, (1999).

[2] Mansuripur, M. *Classical Optics and its Applications*. Cambridge University Press, (2009).

[3] Dennis, M. R. *Topological Singularities in Wave Fields*. PhD dissertation, University of Bristol, (2001).

[4] Dennis, M. R. Polarization singularities in paraxial vector fields: morphology and statistics. *Opt. Commun.* **213**, 201–221 (2002).

[5] Goldstein, H. *Classical Mechanics*. Pearson, 2nd edition edition, (1980).

[6] Hannay, J. H. Cyclic rotations, contractibility and Gauss-Bonnet. *J. Phys. A: Math. Gen.* **31**, L321–L324 (1998).

[7] Weeks, J. R. *The Shape of Space*. Marcel Dekker, New York, (1985).

[8] Henderson, D. W. and Taimina, D. *Experiencing Geometry: in Euclidean, Spherical, and Hyperbolic Spaces*. Prentice Hall, (2001).

[9] Frank, F. C. Orientation mapping. *Metall. Mater. Trans. A* **19**, 403–408 (1988).

[10] Frank, F. C. The conformal neo-eulerian orientation map. *Philos. Mag. A* **65**, 1141–1149 (1992).

[11] Byerly, W. E. *An elementary treatise on Fourier's series and spherical, cylindrical, and ellipsoidal harmonics: with applications to problems in mathematical physics*. Ginn & Co., Boston, (1895).

[12] Bateman, H. *Partial Differential Equations of Mathematical Physics*. Dover, (1944).

[13] Moon, P. and Spencer, D. E. *Field Theory Handbook*. Springer-Verlag Berlin Heidelberg, (1988).

[14] Siegman, A. E. *Lasers*. University Science Books, (1986).

[15] Nye, J. F. *Natural Focusing and Fine Structure of Light: Caustics and Wave Dislocations*. IoP Publishing, (1999).

[16] Dennis, M. R., O'Holleran, K., and Padgett, M. J. Singular Optics: Optical Vortices and Polarization Singularities. *Prog. Optics* **53**, 293–363 (2009).

[17] Dennis, M. R., King, R. P., Jack, B., O'Holleran, K., and Padgett, M. J. Isolated optical vortex knots. *Nat. Phys.* **6**, 118–121 (2010).

[18] Dennis, M. R., Goette, J. B., King, R. P., Morgan, M. A., and Alonso, M. A. Paraxial and nonparaxial polynomial beams and the analytic approach to propagation. *Opt. Lett.* **36**, 4452–4454 (2011).

[19] Bauer, T., Banzer, P., Karimi, E., Orlov, S., Rubano, A., Marrucci, L., Santamato, E., Boyd, R. W., and Leuchs, G. Observation of optical polarization Möbius strips. *Science* **347**, 964–966 (2015).

[20] Larocque, H., Sugic, D., Mortimer, D., Taylor, A. J., Fickler, R., Boyd, R. W., Dennis, M. R., and Karimi, E. Reconstructing the topology of optical polarization knots. *Nat. Phys.* **14**, 1079–1082 (2018).

[21] Sugic, D. *Unravelling the dark focus of light: a study of knotted optical singularities*. PhD dissertation, University of Bristol, (2019).

[22] Otte, E., Rosales-Guzmán, C., Ndagano, B., Denz, C., and Forbes, A. Entanglement beating in free space through spin-orbit coupling. *Light: Sci. & Appl.* **7**, 18007–18009 (2018).

[23] Schulze, C., Flamm, D., Duparré, M., and Forbes, A. Beam-quality measurements using a spatial light modulator. *Opt. Lett.* **37**, 4687–4689 (2012).

[24] Schaefer, B., Collett, E., Smyth, R., Barrett, D., and Fraher, B. Measuring the Stokes polarization parameters. *Am. J. Phys.* **75**, 163–168 (2007).

[25] Otte, E., Alpmann, C., and Denz, C. Higher-order polarization singularitites in tailored vector beams. *J. Opt.* **18**, 074012 (2016).

[26] Alpmann, C., Schlickriede, C., Otte, E., and Denz, C. Dynamic modulation of Poincaré beams. *Sci. Rep.* **7**, 8076 (2017).

[27] Takeda, M., Ina, H., and Kobayashi, S. Fourier-transform method of fringe-pattern analysis for computer-based topography and interferometry. *J. Opt. Soc. Am.* **72**, 156–160 (1982).

[28] Čižmár, T., Mazilu, M., and Dholakia, K. In situ wavefront correction and its application to micromanipulation. *Nat. Photonics* **4**, 388–394 (2010).

[29] Beckley, A. M., Brown, T. G., and Alonso, M. A. Full Poincaré beams. *Opt. Express* **18**, 10777–10785 (2010).

[30] Donati, S., Dominici, L., Dagvadorj, G., Ballarini, D., De Giorgi, M., Bramati, A., Gigli, G., Rubo, Y. G., Szymańska, M. H., and Sanvitto, D. Twist of generalized skyrmions and spin vortices in a polariton superfluid. *Proc. Natl. Acad. Sci. U.S.A.* **113**, 14926–14931 (2016).

[31] Gao, S., Speirits, F. C., Castellucci, F., Franke-Arnold, S., Barnett, S. M., and Götte, J. B. Paraxial skyrmionic beams. *Phys. Rev. A* **102**, 053513 Nov (2020).

[32] Salomaa, M. and Volovik, G. E. Quantized vortices in superfluid He-3. *Rev. Mod. Phys.* **59**, 533–613 (1987).

[33] Leanhardt, A. E., Shin, Y., Kielpinski, D., Pritchard, D. E., and Ketterle, W. Coreless Vortex Formation in a Spinor Bose-Einstein Condensate. *Phys. Rev. Lett.* **90**, 140403 (2003).

[34] Nagaosa, N. and Tokura, Y. Topological properties and dynamics of magnetic skyrmions. *Nat. Nanotechnol.* **8**, 899–911 (2013).

[35] Tsesses, S., Ostrovsky, E., Cohen, K., Gjonaj, B., Lindner, N. H., and Bartal, G. Optical skyrmion lattice in evanescent electromagnetic fields. *Science* **361**, 993–996 (2018).

[36] Du, L., Yang, A., Zayats, A. V., and Yuan, X. Deep-subwavelength features of photonic skyrmions in a confined electromagnetic field with orbital angular momentum. *Nat. Phys.* **15**, 650–654 (2019).

[37] Mermin, N. D. and Ho, T.-L. Circulation and angular momentum in the *A*-phase of superfluid Helium-3. *Phys. Rev. Lett.* **36**, 594–597 (1976).

[38] Bott, R. and Tu, L. W. *Differential Forms in Algebraic Topology*. Springer-Verlag, (1982).

[39] Hatcher, A. *Algebraic Topology*. Cambridge University Press, (2002).

[40] Lee, W., Gheorghe, A. H., Tiurev, K., Ollikainen, T., Möttönen, M., and Hall, D. S. Synthetic electromagnetic knot in a three-dimensional skyrmion. *Sci. Adv.* **4**, eaao3820 (2018).

[41] Shankar, R. Applications of topology to the study of ordered systems. *J. Phys. (France)* **38**, 1405–1412 (1977).

[42] Nakahara, M. *Geometry, Topology and Physics*. Taylor & Francis, 2nd edition edition, (2003).

[43] Manton, N. and Sutcliffe, P. *Topological Solitons*. Cambridge University Press, (2004).

[44] Whitehead, J. H. C. An Expression of Hopf's Invariant as an Integral. *Proc. Natl. Acad. Sci. U.S.A.* **33**, 117–123 (1947).

[45] Moffatt, H. K. and Ricca, R. L. Helicity and the Călugăreanu invariant. *Proc. R. Soc. Lond.* **439**, 411–429 (1992).

[46] Hietarinta, J., Jaykka, J., and Salo, P. Dynamics of vortices and knots in Faddeev's model. *Proceedings of Workshop on Integrable Theories, Solitons and Duality — PoS(unesp2002)* **008**, 017 (2002).

[47] Kedia, H., Foster, D., Dennis, M. R., and Irvine, W. T. M. Weaving Knotted Vector Fields with Tunable Helicity. *Phys. Rev. Lett.* **117**, 274501 (2016).

[48] Sutcliffe, P. Knots in the Skyrme-Faddeev model. *Proc. R. Soc. A* **463**, 3001–3020 (2007).

[49] Berry, M. V. Optical currents. *J. Opt. A: Pure Appl. Opt.* **11**, 094001 (2009).

[50] Dennis, M. R. On the Burgers vector of a wave dislocation. *J. Opt. A* **11**, 094002 (2009).

[51] Volovik, G. E. and Mineev, V. P. Particle-like solitons in superfluid $^3$He phases. *JETP Letters* **46**, 401–404 (1977).

[52] Moffatt, H. K. The degree of knottedness of tangled vortex lines. *J. Fluid Mech.* **35**, 117–129 (1969).

[53] Moffatt, H. K. Helicity and singular structures in fluid dynamics. *Proc. Natl. Acad. Sci. U.S.A.* **111**, 3663–3670 (2014).

[54] Winfree, A. T. and Strogatz, S. H. Singular filaments organize chemical waves in three dimensions III. Knotted waves. *Physica D* **9**, 333–345 (1983).

[55] Ruostekoski, J. Stable particlelike solitons with multiply quantized vortex lines in Bose-Einstein condensates. *Phys. Rev. A* **70**, 041601 (2004).

[56] Berry, M. V. and Dennis, M. R. Knotted and linked phase singularities in monochromatic waves. *Proc. R. Soc. A* **457**, 2251–2263 (2001).

[57] Murakami, J. Volume formulas for a spherical tetrahedron. *Proc. Amer. Math. Soc.* **140**, 3289–3295 (2012).

[58] Jennings, P. R. *Hyperspherical Trigonometry, Related Elliptic Functions and Integrable Systems*. PhD dissertation, The University of Leeds, (2013).